\documentclass[reprint,aps,showpacs,amsmath,twocolumn,groupaddress,letterpaper,floatfix,longbibliography]{revtex4-2}
\usepackage{graphicx,color}
\usepackage{verbatim}
\usepackage{amssymb}   
\usepackage{amsmath}
\usepackage{amsfonts}
\usepackage{mathdots}
\usepackage{hyperref}
\usepackage{epsfig}
\usepackage{xcolor}

\usepackage{braket}
\usepackage{bm}
\begin{document}
\title{ $\varphi_0$-Josephson junction in twisted bilayer graphene induced by a valley-polarized state}
	\author{Ying-Ming Xie$^1$}\thanks{ymxie@ust.hk}
	\author{Dmitri K. Efetov$^2$}
     \author{K. T. Law$^1$} \thanks{phlaw@ust.hk}
        \affiliation{$^1$ Department of Physics, Hong Kong University of Science and Technology, Clear Water Bay, 999077 Hong Kong, China}
     	\affiliation{$^2$ICFO - Institut de Ciencies Fotoniques, The Barcelona Institute of Science and Technology, Castelldefels, Barcelona, 08860, Spain}

\date{\today}
\begin{abstract}
	 Recently, gate-defined Josephson junctions in magic angle twisted bilayer graphene (MATBG) were studied experimentally and highly unconventional Fraunhofer patterns were observed. In this work, we show that an interaction-driven valley-polarized state connecting two superconducting regions of MATBG would give rise to a long-sought-after purely electric controlled  $\varphi_0$-junction in which the two superconductors acquire a finite phase difference $\varphi_0$ in the ground state. We point out that the emergence of the $\varphi_0$-junction stems from the \textcolor{black}{ valley-polarized state which breaks time-reversal symmetry} and trigonal warping effects which break  intravalley inversion symmetry. Importantly, a spatially non-uniform valley polarization order parameter at the junction can explain the key features of the observed unconventional Fraunhofer patterns. Our work explores the novel transport properties of the valley-polarized state, and we suggest that gate-defined MATBG Josephson junctions could realize the first purely electric controlled  $\varphi_0$-junctions. \end{abstract}

\pacs{}

\maketitle

\section{Introduction}

The discovery of correlated insulating states and superconducting states in magic angle twisted bilayer graphene (MATBG) \cite{Bistritzer2011,Cao2018,Cao2018sc} motivated intense studies of moir\'e materials in recent years. The rich symmetry breaking states discovered in MATBG \cite{Liang2018,Noah2018,Isobe2018,Fan2018, FengCheng2018,Cenke2018,LianBiao2019, Jiang2019,Stauber2019,Yankowitz2019, Kerelsky2019,Xie2019,Choi2019,Lu2019,David2019,Young2020,Stepanov2020,Yazdani2020,Vafek2019,Zhida2019,Saito2020,Zondiner2020,Choi2021,Das_Sarma2020,Po2018,Yahui2019,Macdonal2020, Zaletel2020,Zaletel2020_2,Guinea2020,Fuchun2020,Xidai2021, Oreg2021,Caoyuan_2021_nematicity, Nori2020} enable  the creation of novel quantum devices with various quantum phases on a single material platform. Recently, gate-defined Josephson junctions (JJs) were created on MATBG \cite{Rodan-Legrain2021,deVries2021,Diez_Merida2021,Fokert_SQUID_2022} when a non-superconducting (weak-link) region in a superconducting MATBG device was created by local gating. Interestingly,  a highly unconventional Fraunhofer pattern was observed in Ref.~\cite{Diez_Merida2021} when the weak-link region was gated to near half-filling $\nu=-1/2$ filling (two holes per moir\'e unit cell). 

\begin{figure}
	\centering
	\includegraphics[width=1\linewidth]{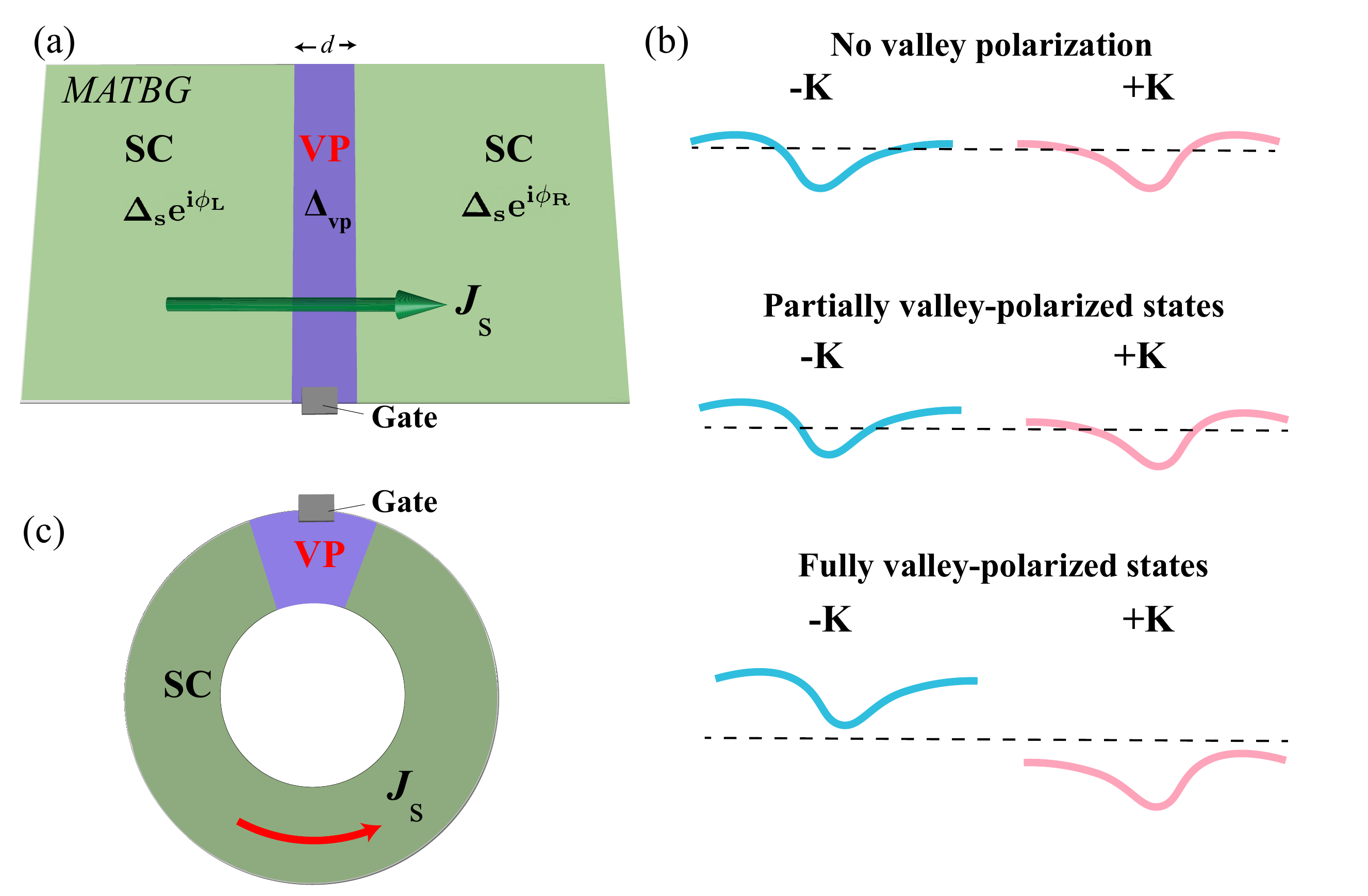}
	\caption{ (a) A schematic plot of a  gate-defined MATBG Josephson junction. The left (right) side of the junction is superconducting with pairing order parameter $\Delta_{\text{s}}e^{i\phi_{\text{L(R)}}}$. The weak-link region has width $d$ and a valley polarization order parameter $\Delta_{\text{vp}}$.  (b) Illustrations of the moir\'e bands at $K$ and $-K$ valleys which are not valley polarized, partially valley-polarized, and fully valley-polarized, respectively \cite{Zaletel2020,Matthew2022}.  The black dashed lines denote the Fermi levels. (c) A schematic plot of a MATBG superconducting ring with a region gated to the valley-polarized state. The $J_s$ (red arrow) represents a spontaneous supercurrent. } 	\label{fig:fig1}
\end{figure}

    The observed unconventional Fraunhofer pattern motivated us to study the Josephson effects in a gate-defined superconductor/valley-polarized state/superconductor (SC/VP/SC) in MATBG, as schematically shown in Fig.~\ref{fig:fig1}(a). \textcolor{black}{As the unconventional Fraunhofer pattern indicates time-reversal and inversion symmetry breaking at the weak-link of the Josephson junction \cite{Diez_Merida2021}, we choose the weak link to be a partially valley-polarized state. In this case, the energy degeneracy  of moir\'e bands of the K and -K valleys is broken due to electron-electron interactions [Fig.~\ref{fig:fig1}(b)]. Such a valley-polarized state is one of the possible energetically favourable states at half-filling from Hartree-Fock calculations \cite{Po2018,Yahui2019,Macdonal2020,Zaletel2020,Zaletel2020_2,Fuchun2020,Guinea2020, Xidai2021, Oreg2021,Matthew2022}, which also satisfies the symmetry requirements of the experiment. The choice of the valley-polarized state as the weak-link in the Josephson junction is further motived by the observation of anomalous Hall effect at half-filling  in the recent experiment, in which the twist angle is slightly away from the magic angle \cite{Matthew2022}. This anomalous Hall effect can also be explained by the partially valley-polarized state.} 
    
    In this work, we show that the current-phase relation induced by the interaction-driven valley-polarized state as the weak link of a Josephson junction is highly unconventional, which has the form $I_\text{s}=I_\text{c} \sin(\phi-\varphi_0)$.  Here, $I_\text{c}$ is the critical current, $\phi=\phi_\text{L}-\phi_{\text{R}}$ is the phase difference of the two superconductors with phases $\phi_\text{L}$ and $\phi_{R}$ respectively. Such Josephson junctions with general $\varphi_0$ are called $\varphi_0$-Josephson junctions ($\varphi_0$-JJs). We further point out that the  valley polarization and the trigonal warping effects are the key ingredients for realizing $\varphi_0$-JJs.  Importantly, a spatially non-uniform valley polarization order parameter at the junction can provide a plausible explanation for the unconventional Fraunhofer patterns observed in the experiment \cite{Diez_Merida2021}. 

 $\varphi_0$-JJs  have important potential device applications, such as superconducting spintronics \cite{Linder2015,Eschrig_2015}, Josephson qubits \cite{Ioffe1999,Yamashita2005,Padurariu2010}, and phase batteries \cite{Strambini2020}. The previously proposed realizations of $\varphi_0$-JJs involve   ferromagnetic materials \cite{Braude2007,Grein2009,Beenakker2009,Enoksen2012,Konschelle2009,JunFeng2010,Bergeret2017} or materials with spin-orbit coupling \cite{ Buzdin2008, Budin2017, Reynoso2008, Martin2009, Nazarov2014,  Tanaka2009,Meyer2015,Bergeret_2015, Bergeret_2015_PRB,PRB2017,Szombati2016,Assouline2019,Mayer2020,Mohammad2021}.  However, experimentally realizations of  $\varphi_0$-JJ were rare and the presence of external magnetic fields was needed \cite{Strambini2020, Szombati2016,Assouline2019,Mayer2020}. \textcolor{black}{This work establishes a new platform of realizing $\varphi_0$-JJs with the interaction-driven valley-polarized state in MATBG.}
 
  
\section{Model for numerical calculation}

First, we introduce a microscopic model which describes a MATBG Josesphon junction as realized experimentally in Ref.~\cite{Diez_Merida2021} and  schematically shown  in Fig.~\ref{fig:fig1}(a).  The relevant moir\'e bands near charge neutrality of MATBG can be captured by an effective two-orbital  tight-binding model on a hexagonal lattice~\cite{Noah2018,Liang2018}, which can be written as :
\begin{align}
	H_0&=\sum_{\braket{ij},\xi\sigma }t_1c_{i\xi\sigma}^{\dagger}c_{j\xi\sigma}+\sum_{\braket{ij}',\xi\sigma}t_{2\xi}c^{\dagger}_{i\xi\sigma}c_{j\xi\sigma}\nonumber\\&+\text{H.c.}-\sum_{i,\xi\sigma}\mu_ic^{\dagger}_{i\xi\sigma}c_{i\xi\sigma}.
\end{align}
Here ,  $\xi$ labels the two $p$-wave-like orbitals $p_x+i\xi p_y$  as a representation of two valleys $\tau=\pm K$ ,  \textcolor{black}{$\sigma=\uparrow/\downarrow$ denotes the spin indices},  $t_1=0.331$ meV and $t_{2\xi}=-0.01+0.097\xi i$ meV denote the  first-nearest neighbor and the fifth-nearest neighbor hopping. Note that the imaginary part of $t_{2\xi}$ describes the warping effects. Moreover, the spatial dependent chemical potential is denoted by $\mu$ which is chosen such that the filling factor $\nu$ satisfies $-1<\nu<-1/2$ for the superconducting part of the junction and $\nu\approx-1/2$ at the weak-link region \cite{Diez_Merida2021}. As shown in Ref.~\cite{Noah2018,Liang2018},  $H_0$ captures the symmetries of the moir\'e bands of MATBG.

To include the effects of interactions, we introduce the superconducting order parameter on the left (L) and right (R) sides of the Josephson junction and the valley polarization order parameter to the weak link. The resulting effective tight-binding Hamiltonian is:
\begin{eqnarray}
	H_{eff}&=H_0+\sum_{i\in \text{(L,R)}, \xi}(\Delta_\text{s}e^{i\phi_{\text{L(R)}}}c^{\dagger}_{i\xi\uparrow}c^{\dagger}_{i-\xi\downarrow}+\text{H.c.})\nonumber\\&+\sum_{i\in \text{WL},\xi\sigma}\Delta_{\text{vp}}c^{\dagger}_{i\xi\sigma}(\tau_z)_{\xi\xi'}c_{i\xi'\sigma}. \label{effective_TB}
\end{eqnarray}
Here, the second term characterizes the pairing potential on the left and right  side of the Josephson junction with phases $\phi_{\text{L}}$ and $\phi_{\text{R}}$ respectively. To be specific, we set the spin-singlet pairing  \cite{LianBiao2019,Fengcheng2020} amplitude  $\Delta_{\text{s}}=0.1$ meV according to  the experiments \cite{Cao2018sc, Diez_Merida2021}, which is roughly one order smaller than the moir\'e band width. It is important to note that other time-reversal invariant unconventional pairings have been proposed in MATBG \cite{Caoyuan_2021_nematicity,Oh2021}. For simplicity,  a conventional spin-singlet pairing order parameter is assumed in the main text.  The conclusions obtained here  are still valid even if we assume other momentum independent pairings which involve both the spin and valley degrees of freedom of MATBG (Appendix E). The temperature effects on the pairing can be included by setting $\Delta_{\text{s}} (T)=\Delta_{\text{s}}\tanh(1.74\sqrt{(T_c-T)/T})$ ($T_c$ is the superconducting critical temperature) \cite{tinkham2004}.

On the other hand, the third term with the Pauli matrix $\tau_z$ characterizes the valley polarization  in the weak-link (WL) region with  valley-polarization order parameter $\Delta_{\text{vp}}$. The order parameter $\Delta_{\text{vp}}$ can be seen from the Hamiltonian with Coulomb interactions under the Hartree-Fock mean-field approximation [see Appendix A]. More details about the tight-binding model can be found in the Appendix C.

\begin{figure}
	\centering
	\includegraphics[width=1\linewidth]{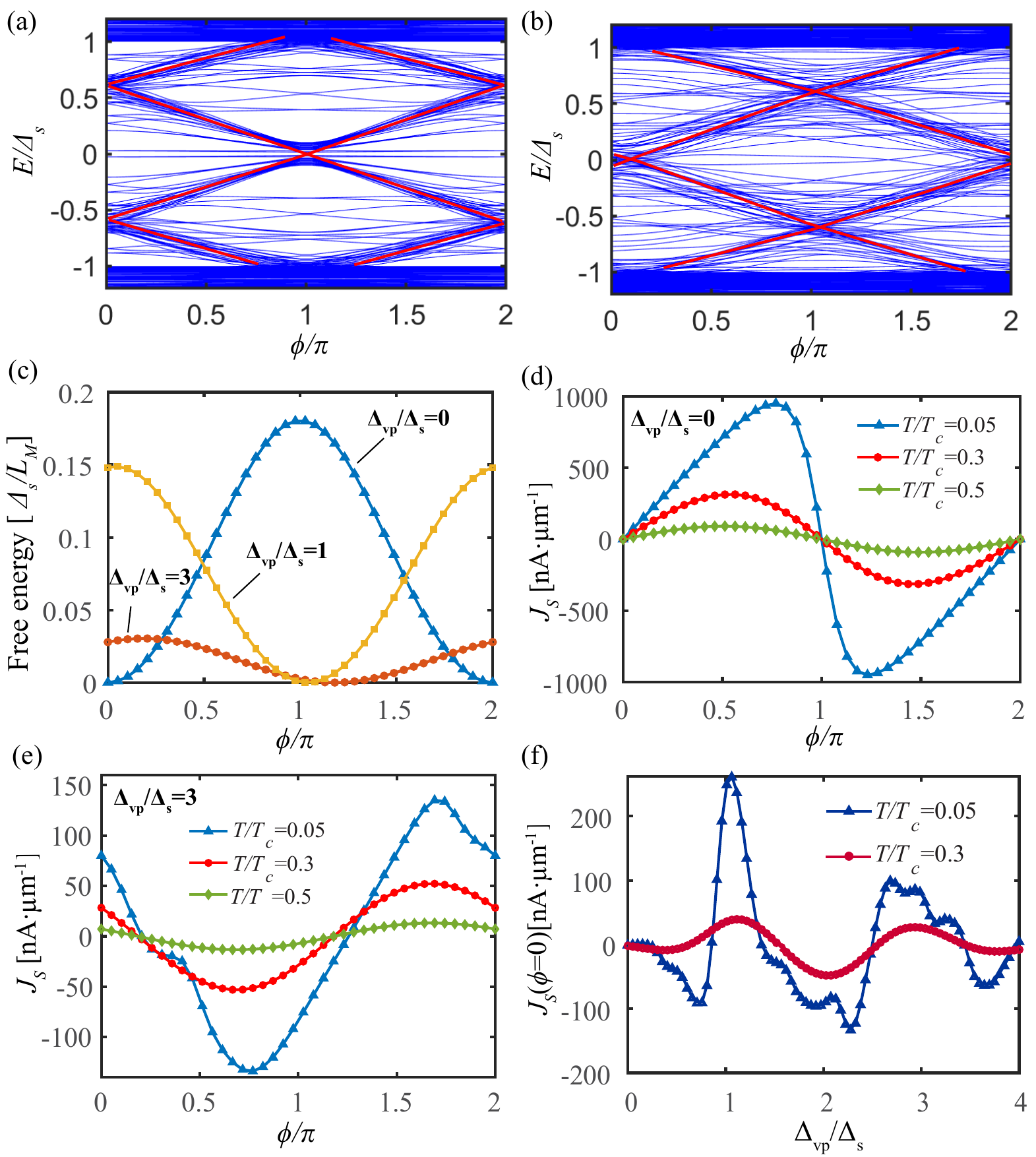}
	\caption{(a) and (b) The energy levels of a MATBG Josephson junction versus the phase difference $\phi$ in the cases of no valley polarization ($\Delta_{\text{vp}}/\Delta_\text{s}=0$) and with a valley polarization $\Delta_{\text{vp}}/\Delta_\text{s}=1$, respectively, where red lines highlight the positions of Anreev bound states with large slopes. The width of the  junction is $W_J=10 L_M$ and length is $d=10L_M/\sqrt{3}$. The filling of weak-link region and the superconducting region are  $\nu\approx -0.5$ and $\nu\approx -0.6$, respectively.  (c) The landscape of the free energy $F_J(\phi)$ (in units of $\Delta_{\text{s}}/L_M$)   with $\Delta_{\text{vp}}/\Delta_\text{s}=0,1,3$. The temperature is fixed at $T/T_c=0.3$.  (d) and (e) The supercurrent density (in the unit of nA$\cdot$ $\mu$m$^{-1}$) versus $\phi$ at various temperatures in the cases of $\Delta_{\text{vp}}/\Delta_\text{s}=0$ and  $\Delta_{\text{vp}}/\Delta_\text{s}=3$, respectively. \textcolor{black}{The total supercurrent across the junction  would be given by $J_sW_J$}.  (f) The anomalous supercurrent $J_s(\phi=0)$ versus the valley polarization strength $\Delta_{\text{vp}}$ at $T/T_c=0.05$ and $T/T_c=0.3$.  }
	\label{fig:fig2}
\end{figure}

\section{Unconventional Josephson junction induced by the valley-polarized state} 

To study the properties of the gate-defined MATBG Josephson junction, we first calculate the energy dispersion as a function of phase difference $\phi$ of the junction which is described by $H_{eff}$. Here, we set the length of the non-superconducting part of the junction be $d=10L_M/\sqrt{3}$ \cite{Diez_Merida2021}  ($L_M\approx$14 nm is the moire lattice constant), and set the filling $\nu$ to be close to half-filling. To match the experimental situation in which the junction resistance is much smaller than the quantized resistance $h/e^2$ \cite{Diez_Merida2021}, we set the weak-link regime to be partially valley-polarized \cite{Zaletel2020,Matthew2022} such that the weak-link section is metallic as schematically illustrated in Fig.~\ref{fig:fig1}(b). The case of fully valley-polarized topological state is studied in the Appendix C. 

Figures \ref{fig:fig2}.~(a) and (b)  show a typical energy spectrum of the MATBG Josephson junction as a function of the phase difference $\phi=\phi_\text{L}-\phi_\text{R}$, obtained by diagonalizing the junction Hamiltonian $H_{eff}$ with $\Delta_{\text{vp}}/\Delta_{\text{s}}=0$ and $\Delta_{\text{vp}}/\Delta_{\text{s}}=1$, respectively. As expected, there is a large number of Andreev bound states within the superconducting gap. The energy-phase relations of a few Andreev bound states with large slopes are highlighted by red solid lines in  Figs.~\ref{fig:fig2}(a) and~\ref{fig:fig2}(b). It can be seen that the in-gap Andreev bound states with large slopes $\frac{\partial E}{\partial\phi} $  contributing mostly to the supercurrent exhibit a phase shift which is close to (but not equal to) $\pi$ when the valley polarization $\Delta_{\text{vp}}/\Delta_{\text{s}}=1$. This phase shift gives the first indication that the valley polarization has nontrivial effects on the Josephson junction. 
 
To study the ground  state of the Josephson junction, we calculate the free energy as 
\begin{equation}
F(\phi)=-k_BT\sum_{n}\ln (1+e^{-E_n(\phi)/k_BT}),
\end{equation}
 where  $T$ is the temperature,  the energy of the states $E_n(\phi)$ is obtained by diagonalizing  the Hamiltonian $H_{eff} (\phi)$. For convenience sake, we define the  free energy of the Josephson junction per unit width to be $F_J(\phi)= W_J^{-1}(F(\phi)-\text{min}[F(\phi)])$, where $W_J$ is the width of the junction. Therefore, the phase difference of the two superconductors at the ground state is determined by $\phi_0$ such that $F_J(\phi_0)=0$.  In Fig.~\ref{fig:fig2}(c), we plot the free energy landscapes $F_J(\phi)$ with temperature  $T=0.3T_c$ at various valley polarization strengths ($\Delta_{\text{vp}}/\Delta_{\text{s}}=0,1,3$).  As expected, without valley polarization, the junction is conventional so that the ground state appears at $\phi=0$.  Interestingly, the ground state of the Josephson junction can appear at a finite $\phi$ in the presence of valley polarization.  For example, in the case of $ \Delta_{\text{vp}}/\Delta_{\text{s}}=1$, the ground state with $F_J(\phi)=0$ appears at a phase difference  close to (but not equal to) $\pi$. For a larger $\Delta$ such that $ \Delta_{\text{vp}}/\Delta_{\text{s}}=3$, the ground state appears at a phase further away from $\pi$. 

To show the effect of valley polarization on the current-phase relation, the supercurrent density $J_s$   (in unit of nA$\cdot$ $\mu$m$^{-1}$) as a function of $\phi$ is depicted in Fig.~\ref{fig:fig2}(d)  for the case without valley polarization ($\Delta_{\text{vp}}/\Delta_{\text{s}}=0$) and in Fig.~\ref{fig:fig2}(e) for the case with valley polarization ($\Delta_{\text{vp}}/\Delta_{\text{s}}=3$). Here, the supercurrent density is obtained from the free energy of the Josephson junction as  $J_s=\frac{2e}{\hbar}\frac{\partial F_J(\phi)}{\partial \phi}$. Without valley polarization, the  junction has conventional  current-phase relation at both the low and high temperature regimes \cite{Beenakker1991,Beenakker1992}. However, in the case with finite valley polarization, the supercurrent can either exhibit a sign change or even display a generic phase shift [see Fig.~\ref{fig:fig2}(e)]. In particular, it can be seen that the curves with higher temperature [the red and green lines in Fig.~\ref{fig:fig2}(e)]  follow a standard $\varphi_0$-JJ current-phase relation of $J_\text{s}=J_\text{c} \sin(\phi-\varphi_0)$.  Our calculation thus clearly shows that the valley polarization can result in $\varphi_0$-JJs in MATBG. 

 One important consequence of a $\varphi_0$-JJ is that there is a supercurrent even at zero phase difference ($\phi=0$), called anomalous supercurrent \cite{Reynoso2008,Martin2009}.  The anomalous supercurrent density $J_s(\phi=0)$ for $\Delta_{\text{vp}}/\Delta_{\text{s}}=3$  can  be  seen in  Fig.~\ref{fig:fig2}(e). The $J_s(\phi=0)$  as a function of valley polarization strength $\Delta_{\text{vp}}$  at various temperatures is shown in Fig.~\ref{fig:fig2}(f). We find that the anomalous supercurrent is generally finite with valley polarization. Moreover, when  $\Delta_{\text{vp}}\gg \Delta_{\text{s}}$, the anomalous current density at the low temperature range can be as large as tens of nA$\cdot$ $\mu$m$^{-1}$. As depicted in Fig.~\ref{fig:fig1}(c), we expect to see an anomalous current in a ring geometry when part of the superconducting ring is gated to the valley-polarized state. It is also important to note that  unlike previously studied $\varphi_0$-JJs,  the MATBG $\varphi_0$-JJs  do not  involve ferromagnetism or spin-orbit coupling, which calls for a new understanding about the underlying mechanism for the formation of $\varphi_0$-JJs in the MATBG. 

\section{Underlying mechanism for  $\varphi_0$-JJs in MATBG}

Next, based on the scattering matrix method \cite{Beenakker1991,Beenakker1992}, we show analytically that the valley-polarization and the warping effects of moir\'e bands are crucial in realizing a $\varphi_0$-JJ. At the junction, the states can be labelled by the transverse momentum $k_y$. For illustration, we demonstrate how the Andreev bound state associated with the $k_y=0$ mode (normal incident states),  is affected by valley polarization and the warping terms. The  1D Hamiltonian associated with the $k_y=0$ mode can be written as
$H_{1\text{D}}=\sum_{\tau\alpha}\int dx\Psi^{\dagger}_{\tau\alpha}(x)\hat{H}_{\tau\alpha}(x)\Psi_{\tau\alpha}(x)$. Here, $\tau=+/-$ labels the valley index, $\alpha=+/-$ labels the incoming/outgoing normal states near Fermi energy,  $\Psi_{\tau\nu}=(\psi_{\tau\alpha}(x),\psi^{\dagger}_{-\tau,-\alpha}(x))^{T}$ denotes the  Nambu basis, and 
\begin{equation}
	\hat{H}_{\tau\alpha}(x)=\begin{pmatrix}
	H_{N,\tau\alpha}(x)+\Delta_{\text{vp}}(x)\tau&\Delta_\text{s}(x)\\
		\Delta_\text{s}(x)&	H^*_{N,-\tau-\alpha}(x)+\Delta_{\text{vp}}(x)\tau
	\end{pmatrix}.\label{Ham_junction}
\end{equation}
Here, the linearized single-particle Hamiltonian $H_{N,\tau\alpha}(x)=-i\alpha \hbar v_{f,\tau\alpha}(x)\partial_x$,  the longitudinal  Fermi velocity along the current direction is given by $v_{f,\tau\alpha}$ such that $v_{f,\tau\alpha}(x)=v_{\text{s},\tau\alpha}[\Theta(x)+\Theta(x-d)]+v_{\text{vp},\tau\alpha}\Theta(x)\Theta(d-x)$, where $v_{\text{s},\tau\alpha}$ and $v_{\text{vp},\tau\alpha}$ are the Fermi velocities for the superconducting region and the valley-polarized weak-link region, respectively.  Notably,  the warping term which breaks the intravalley inversion symmetry could lead to $v_{\text{vp},\tau\alpha}\neq v_{\text{vp},\tau-\alpha}$.  The superconducting pairing potential is written as $\Delta_\text{s}(x)=\Delta_\text{s}(e^{i\frac{\phi}{2}}\Theta(-x)+e^{-i\frac{\phi}{2}}\Theta(x-d))$,  and the valley-polarized order parameter is $\Delta_{\text{vp}}(x)=\Delta_{\text{vp}}\Theta(x)\Theta(d-x)$.

With the effective one-dimensional Hamiltonian $H_{1\text{D}}$, we can solve the energies of the Andreev bound states $\epsilon_\tau$ analytically ($\tau$ is a good quantum number), which are given by [for more details see Appendix B]
\begin{equation}	\cos(2\beta-\frac{2(\epsilon_{\tau}-\tau\Delta_{\text{vp}})}{E_T})=\cos(\phi+\frac{\epsilon_{\tau}-\tau\Delta_{\text{vp}}}{\tau E_A}).\label{Andreev_bound}
	\end{equation}

	
Here, $\beta(\epsilon_\tau)=\arccos\frac{\epsilon_\tau}{\Delta_{\text{s}}}$, $E_{T}=\hbar \bar{v}_{\text{vp}}/d$ is the Thouless energy, $E_{A}=\hbar \delta \bar{v}_{\text{vp}}/d$ is an energy scale that reflects the intravalley asymmetry induced by the warping term, where  $\bar{v}_{\text{vp}}$ and $\delta \bar{v}_{\text{vp}}$ are defined by $\bar{v}_{\text{vp}}=4(\sum_{\tau\nu}v_{\text{vp},\tau\alpha}^{-1})^{-1}$ and  $\delta \bar{v}_{\text{vp}}=2(v_{\text{vp},++}^{-1}+v_{\text{vp},--}^{-1}-v_{\text{vp},+-}^{-1}-v_{\text{vp},-+}^{-1})^{-1}$. 

\begin{figure}
	\centering
	\includegraphics[width=1\linewidth]{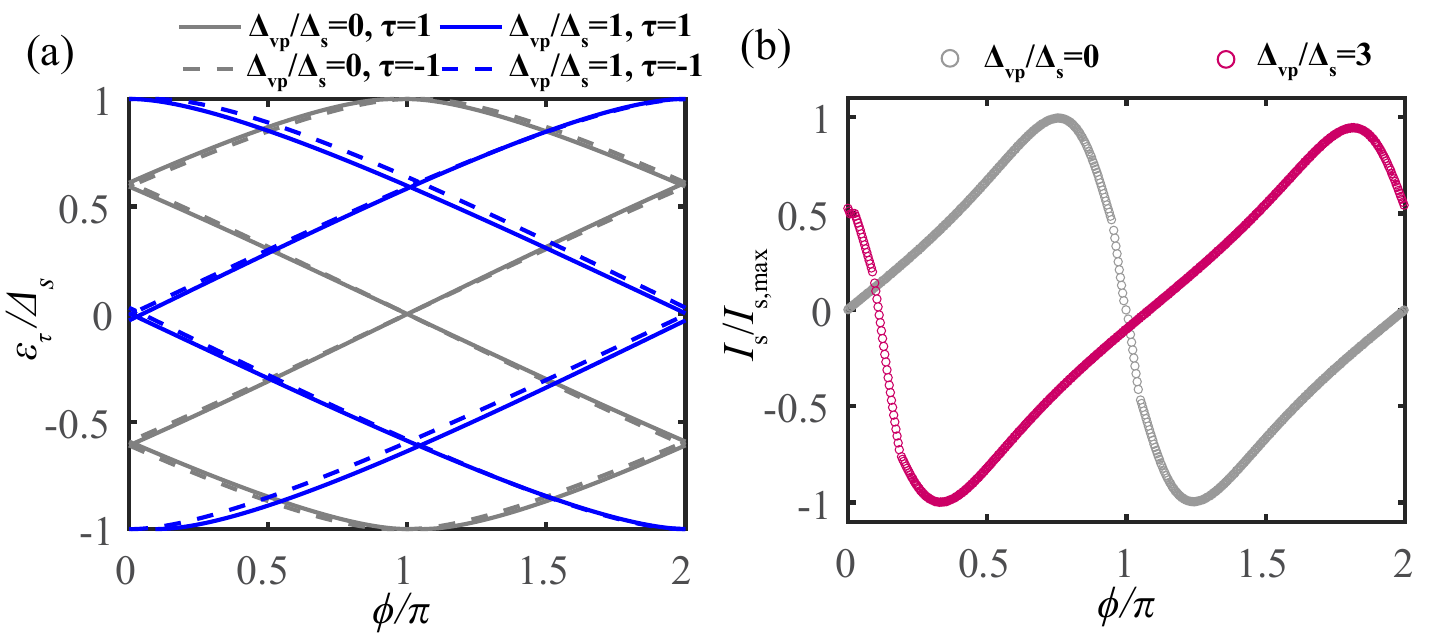}
	\caption{(a) The energies of the Andreev bound states $\epsilon_{\tau}$ versus the phase $\phi$ obtained from Eq.~(\ref{Andreev_bound}) with $\Delta_{\text{vp}}/\Delta_{\text{s}}=0$ (gray lines) and  $\Delta_{\text{vp}}/\Delta_{\text{s}}=1$ (blue lines), where $\tau=\pm 1$ denotes the valley index. (b) The Josephson current $I_s$ (normalized by its maximal value) versus $\phi$ calculated with the Andreev bound states given by Eq.~(\ref{Andreev_bound})  with $\Delta_{\text{vp}}/\Delta_{\text{s}}=0$ (gray dots) and $\Delta_{\text{vp}}/\Delta_{\text{s}}=3$  (red dots), respectively. The two other energy scales are given by $E_T=0.65\Delta_{\text{s}}$ and $E_A=12\Delta_{\text{s}}$, respectively. }
	\label{fig:fig3}
\end{figure}

 Importantly, many features of the numerical results as shown in Fig.~\ref{fig:fig2} can be captured by Eq.~(\ref{Andreev_bound}). For example, we can calculate the Andreev bound state energies associated with  the $\tau=+/-$ valleys by solving $\epsilon_{\tau}(\phi)$  from  Eq.~(\ref{Andreev_bound}) at $\Delta_{\text{vp}}/\Delta_{\text{s}}=0$  and $\Delta_{\text{vp}}/\Delta_{\text{s}}=1$ [see the gray lines and blue lines in Fig.~\ref{fig:fig3}(a), respectively]. Note that in  Fig.~\ref{fig:fig3}(a), the valley degeneracy of Andreev bound states are lifted by the warping term and valley polarization. With the bound state energies $\epsilon_{\tau}$, it is straightforward to obtain the supercurrent $I_s(\phi)$ by adopting the relation
\begin{equation}
	I_\text{s}(\phi)=-\frac{2e}{\hbar}\sum_{\tau}\sum_{\epsilon_{\tau}>0}\tanh(\frac{\epsilon_\tau}{2k_BT})\frac{\partial \epsilon_\tau}{\partial \phi}.
\end{equation}
As an illustration, we plot the calculated $I_\text{s}(\phi)$ at the low temperature limit in the cases of $\Delta_{\text{vp}}/\Delta_{\text{s}}=0$  and  $\Delta_{\text{vp}}/\Delta_{\text{s}}=3$ [see Fig.\ref{fig:fig3}~(b)]. Notably, the features  are in agreement with the ones shown in Figs.~\ref{fig:fig2}(d) and~\ref{fig:fig2}(e).

\begin{figure}
	\centering
	\includegraphics[width=0.9\linewidth]{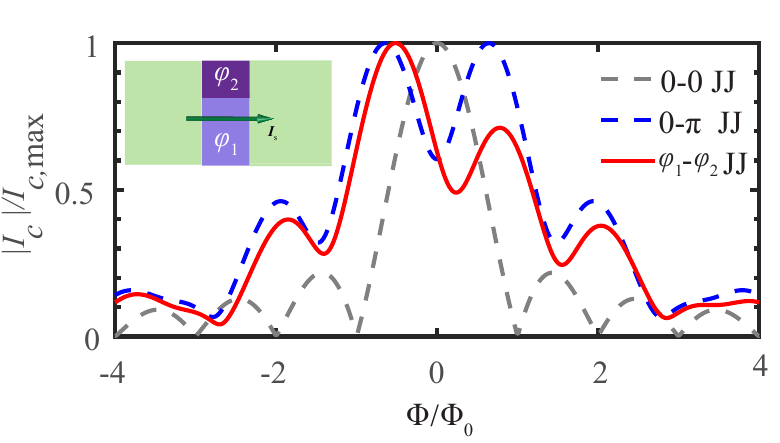}
	\caption{The Fraunhofer patterns for different types of Josephson junctions. The critical current $I_c$ (normalized by its maximal value) versus the magnetic flux $\Phi$ passing through the junction (in units of $\Phi_0=h/2e$) for the 0-0 (gray), $0$-$\pi$  (blue) and $\varphi_1$-$\varphi_2$ (red) Josesphson junctions respectively. The inset illustrates a simple two-domain model with different valley polarization induced phase shifts $\varphi_1$ and $\varphi_2$. The Fraunhofer pattern of the $\varphi_1$-$\varphi_2$ junction matches the experimental observations well with $\varphi_1=0.2$,  $\varphi_2=\pi+0.8$. The widths  of the $\varphi_1$ and the $\varphi_2$ sections are chosen to have the ratio $W_{J1}/W_{J2} \approx 2.3$. }
	\label{fig:fig4}
\end{figure}

In the short junction and at  the high temperature limit, we can obtain an analytical form for  the Josephson current:
\begin{equation}
	I_\text{s}(\phi)\approx \frac{e\Delta_\text{s}^2}{2\hbar k_BT}\cos(\frac{2\Delta_{\text{vp}}}{E_T})\sin(\phi-\varphi_0).\label{analytical}
\end{equation}
It can be seen that $I_s(\phi)$ indeed significantly differs from the conventional form given by $I_s(\phi) \propto \sin\phi$.  Specifically, the supercurrent exhibits a phase shift  $\varphi_0=\Delta_{\text{vp}}/E_A$, which is determined by both the valley polarization and Fermi velocity asymmetry induced by warping effects  ($\varphi_0=0$ if there is no asymmetry as $E_A\rightarrow \infty$). Remarkably, the factor $\cos(\frac{2\Delta_{\text{vp}}}{E_T})$  indicates that the supercurrent oscillates periodically as a function of $\Delta_{\text{vp}}$, being consistent with the numerical result  in Fig.~\ref{fig:fig2}(f). It is worth noting that  the results are similar to Eq.~\ref{analytical} for modes with small transverse momentum $k_y$ as well. 

\section{Unconventional Fraunhofer pattern}

In practice, the sample inhomogeneity and the formations of valley polarization domain walls may lead to spatially non-uniform $\Delta_{\text{vp}}$ inside the weak-link region \cite{Zeldov2022}, which can affect the transport properties of the $\varphi_0$-JJ. As an illustration, we calculate the Fraunhofer pattern for a simple   geometry with two  valley polarization  domains (see the inset of Fig.~\ref{fig:fig4}), in which each domain generates a phase difference of $\varphi_1$ and $\varphi_2$ respectively at the junction. Such a $\varphi_1$-$\varphi_2$ junction is a generalization of the previously studied $0$-$\pi$ junctions \cite{Weides2006,Frolov2006, Kemmler2010}.  Here, we plot the resulting Fraunhofer patterns in Fig.~\ref{fig:fig4} and the details can be found in the Appendix D. Interestingly,  the Fraunhofer pattern in the $\varphi_1$-$\varphi_2$ junction captures the main features found in the recent experiment \cite{Diez_Merida2021} which exhibits a shift in the central peak, a large asymmetry with respect to the central peak, and a non-vanishing critical current as a function of magnetic fields. These features are not naturally expected by conventional Josephson junctions nor $0$-$\pi$ Josephson junctions \cite{Weides2006, Frolov2006, Kemmler2010}. It is worth noting that although our study provides a plausible mechanism for the unconventional Fraunhofer  pattern  seen in the experiment based on $\varphi_0$-JJs, we cannot exclude other ways of generating such Fraunhofer  patterns. To directly view the $\varphi_0$-JJs in the experiment, it would be more straightforward by using the SQUID structures as previous experiments \cite{Szombati2016,Assouline2019}, which are discussed in detail in Appendix D.

\section{ Discussion}

It is important to note that our model  exhibits some similarities between the $\varphi_0$-JJ model induced by Rashba spin-orbit coupling and exchange field \cite{Buzdin2008,Bergeret_2015} by regarding the valley as a pseudospin. Specifically,  the trigonal  warping term and valley polarization can play the role of spin-orbit coupling and spin polarization, respectively. The detailed mapping is illustrated in the Appendix F. This mapping provides  a good insight about the appearance of the $\varphi_0$ Josephson junction in the MATBG platform, although in which the spin-orbit  coupling is  negligible. We expect $\varphi_0$ Josephson is only an example, while other profound physics that  typically arises from the interplay of spin-orbit  coupling and ferromagnetism can also be explored in the MATBG according to our framework.

 In the main text, we focus on the Josephson junction in the ballistic limit. On the other hand, intervalley backscatterings can couple the two valleys and effectively weaken the valley polarization and reduce $\varphi_0$, similar to the spin-relaxation effects in $\varphi_0$-JJ with spin-orbit coupling \cite{Bergeret_2015, Bergeret_2015_PRB, Luyao_2019}. However, as long as the valley polarization is finite, we still expect a $\varphi_0$-JJ. Furthermore, our work can be easily extended to study the unconventional Josephson effects mediated by valley-polarized states in other moir\'e materials/superconductor heterostructures.

 
 \section*{Acknowledgments}
  The authors thank the discussions with Kin Fai Mak, Jaime Diez-Merida and Adrian Po. K.T.L. acknowledges the support of the Ministry of Science and Technology, China and the Hong Kong Research Grant Council through  Grants No. 2020YFA0309600, No. RFS2021-6S03,No.  C6025-19G,  No.  AoE/P-701/20,  No.  16310520,  No.16310219,  No.  16307622,  and  No.  16309718. Y.M.X.  acknowledges  the  support  of  Hong Kong Research Grant Council through Grant No. PDFS2223-6S01.

	\begin{appendix}
		\section{Trigonal warping effects of moir\'e bands and  valley-polarized states}

		\subsection{Moir\'e bands of MATBG and trigonal warping effects}

\begin{figure}
			\centering
			\includegraphics[width=1\linewidth]{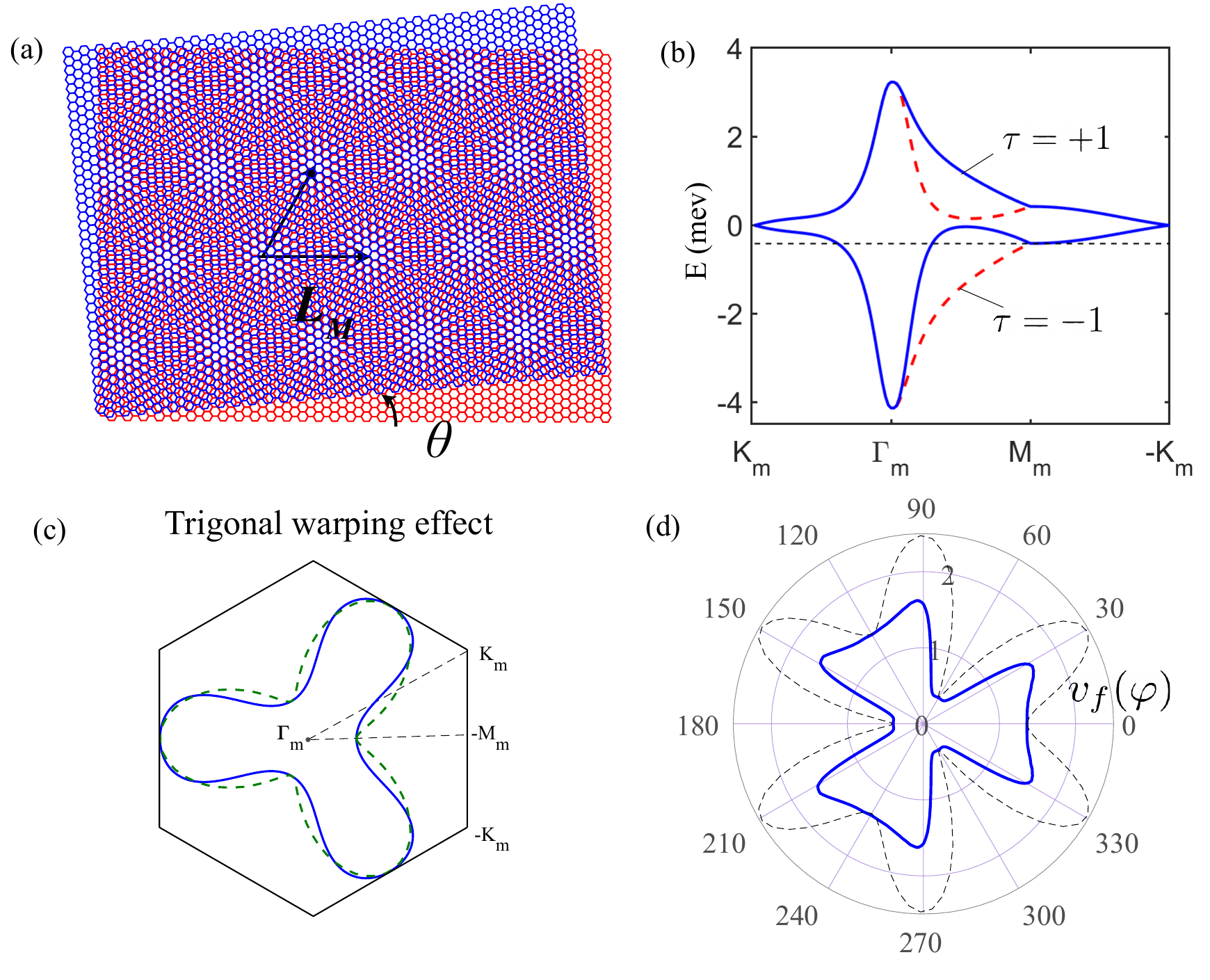}
			\caption{(a) A schematic plot of the  twisted bilayer graphene moir\'e superlattice ($L_M$ is the lattice constant), which is formed using two layers of graphene with a small twisted angle $\theta$.  (b) The lowest moir\'e bands  of twisted bilayer graphene near charge neutrality points with the magic twist angle $\theta=1.05^\circ$. The blue/red bands are  from +K/-K valley ($\tau=+1/\tau=-1$). Note that we have shifted the charge neutrality points to be at zero energy. (c) The  Fermi contour (in blue) near half-filling [the black dashed line in (a)]. The green dashed line is plotted with $k_f(\varphi)=a+b\cos3\varphi$. The black solid lines label the moir\'e Brillouin zone.  (d) The blue line shows the anisotropic Fermi velocity along the  radical direction $v_f(\varphi)$. The black lines are a plot of $v_f(\varphi)$ estimated using $k_f(\varphi)$. The radius is in the unit of $\text{meV}/\hbar L_M^{-1}\approx 2 \times 10^4$ m/s. }
			\label{fig:figs1}
		\end{figure}

		In the main text,  the trigonal warping impact of moir\'e bands and its important effects on creating  $\varphi_0$-Josephson junctions (JJs) in twisted bilayer graphene is highlighted. Here, we present the details of showing the trigonal warping effects using the continuum model of magic angle twisted bilayer graphene (MATBG) (which is depicted in Fig.~\ref{fig:figs1}(a)).   The continuum model of MATBG (c.f.~\cite{Bistritzer2011,Liang2018}) can be written as
		
		\begin{equation}\label{continuum}
			H_{\tau}(\mathbf{r})=\begin{pmatrix}
				H_{b}(\mathbf{r})&T(\mathbf{r})\\
				T^{\dagger}(\mathbf{r})&H_t(\mathbf{r})
			\end{pmatrix}.
		\end{equation}
		Here, the intra-layer moir\'e Hamiltonian is
		\begin{equation}
			H_{l}=-\hbar vR(\theta)[(\hat{\mathbf{k}}-\mathbf{K}_{\tau}^{(l)})\cdot(\tau\sigma_x,\sigma_y)]R^{\dagger}(\theta),
		\end{equation}
		where $\hbar v/a=2.1354$ eV. Here, $l=t/b$ and $\tau=\pm$ label the top/bottom layers and $\pm$ valleys respectively. The twist angle is denoted as $\theta$, $\sigma_j$ represent the Pauli matrices defined in the AB sublattice space, $\mathbf{K}_{\tau}^{(l)}$ labels the Dirac point at valley $\tau$ of the $l$-layer, and the rotational operator $R(\theta)=\text{diag}(e^{-il\frac{\theta}{2}},e^{il\frac{\theta}{2}})$. The interlayer Hamiltonian  $T(\mathbf{r})$ can be written  as
		\begin{align}
			T(\mathbf{r})&=\begin{pmatrix}
				u&u'\\
				u'&u
			\end{pmatrix}+\begin{pmatrix}
				u&u'e^{-i\omega_{\tau}}\\
				u'e^{i\omega_{\tau}}&u
			\end{pmatrix}e^{-i\tau \mathbf{G}_{2}^M\cdot\mathbf{r}}\nonumber\\
		&+\begin{pmatrix}
				u&u'e^{i\omega_{\tau}}\\
				u'e^{-i\omega_{\tau}}&u
			\end{pmatrix}e^{-i\tau \mathbf{G}_{3}^M\cdot\mathbf{r}}.
		\end{align}
		Here, $\omega_{\tau}=\frac{2\pi}{3}\tau$, $\mathbf{G}^M_i=\frac{4\pi}{\sqrt{3}L_M}(\cos\frac{(i-1)\pi}{3},\sin \frac{(i-1)\pi}{3})$ with the moir\'e unit length $L_M=a/\sin\theta\sim 14$ nm, and we adopt $u=0.0797$ eV, $u'=0.0975$ eV according to Ref.~\cite{Liang2018}. The moir\'e bands can be obtained by diagonalizing the continuum Hamiltonian using the plane wave basis $\psi_{\mathbf{k}}(\mathbf{r})=\sum_{\mathbf{G}}C_{\mathbf{G}}e^{i(\mathbf{k}+\mathbf{G})\cdot\mathbf{r}}$ with $G=n_1\mathbf{G}^M_2+n_2\mathbf{G}^M_3$, where $n_1$, $n_2$ are integers.
		{\tiny }
		Fig.~\ref{fig:figs1}(b) show the band structure of the lowest moir\'e bands near the charge neutrality point of MATBG \cite{Liang2018}. 
		
		To highlight the trigonal warping features of the moir\'e bands, a Fermi energy contour near half-filling of the $\tau=+1$ valley (the black dashed line in Fig.~\ref{fig:figs1}(b)) is plotted in Fig.~\ref{fig:figs1}(c) (blue line). We can denote the warped Fermi energy contour as $k_f(\varphi)$. As $k_f(\varphi)=k_f(\varphi+\frac{2\pi}{3})$ due to the $C_3$ symmetry and an emergent $C_{2x}$ symmetry in each valley such that $k_f(\varphi)=k_f(-\varphi)$. To the lowest order in $\varphi$, we can expand $k_f(\varphi)\approx a+b\cos3\varphi$. By using $a=2.315$, $b=-1.299$ (in unit of $L_{M}^{-1}$), we find that $k_f(\varphi)$ can approximately fit the warped Fermi energy contour of the continuum model. The Fermi energy contour at the $\tau=-1$ valley can be obtained by a time-reversal operation. 
		
		Such prominent  warping behaviour is a direct consequence of the narrow bandwidth and the constraint of $D_3$  point group symmetry of MATBG  \cite{Noah2018,Liang2018}. Here, we present the symmetry transformation properties under the generators of $D_3$ which include a three-fold rotation along the $z$-axis and a two-fold rotation along the $y$-axis. Note that the moir\'e bands are assumed to be decoupled in the valley space so that only the terms that involve $\tau_0$ and $\tau_z$ are allowed. Without loss of generality, we consider the Fermi energy cuts the lower branch of the moir\'e bands only as shown in  Fig.~\ref{fig:figs1}(b). In this case, we can construct a simple symmetry-invariant continuum model near $\Gamma_m$ point as 
		\begin{equation}
			H_{\text{eff}}=\lambda_0(k_x^2+k_y^2)+\lambda_1 k_x(k_x^2-3k_y^2)\tau_z-\mu.\label{Eq_warp}
		\end{equation}
		Here, the first term is the kinetic energy term,  $\mu$ denotes the chemical potential term, and the second term is the warping term which is opposite at the opposite valley to preserve the time-reversal symmetry $T=\tau_x K$ and $C_{2y}=\tau_x$ symmetry. The presence of the warping term breaks the intra-valley inversion symmety as $I_0H_{\text{eff}}(\mathbf{k})I_0^{-1}\neq H_{eff}(-\mathbf{k})$, where under $I_0:\tau \mapsto \tau, \mathbf{k}\mapsto -\mathbf{k}$. As emphasized in the main text, the breaking of intra-valley inversion symmetry together with the valley polarization enables the generation of $\varphi_0$-Josephson effect in MATBG even in the absence of the spin-orbit coupling.

		One of the important consequences of the warping effects is to enable the velocity of incoming and outgoing states in the junction to be asymmetric, which plays a crucial role in creating a nontrivial $\varphi_0$ as as shown in later sections. To show the asymmetry of the Fermi velocity in the moir\'e bands, we plot the angular dependence of Fermi velocity $v_f(\varphi)$ (see the blue line in Fig.~\ref{fig:figs1}(d), in unit of $\text{meV}/\hbar L_M^{-1}$), which is defined as $v_f(\varphi)=\sqrt{v_x(k_f(\varphi))^2+v_y(k_f(\varphi))^2}$, $k_f(\varphi)$ is the Fermi momentum contour as shown in Fig.~\ref{fig:figs1}(c). In other words, $v_f(\varphi)$ is the Fermi velocity along the radical direction at each $\varphi$. For normal incident states, we can estimate the asymmetry of the Fermi velocity is given by
		$v_f(\varphi=0)-v_f(\varphi=\pi)\approx	 \text{meV}/\hbar L_M^{-1}\approx 2 \times 10^4 \text{ m/s}.$
		Note that the Fermi velocity is two orders smaller than that of monolayer graphene due to the formation of flat bands under moir\'e superlattice potential.

		To highlight the anisotropy of Fermi velocity  induced by the warping term in Eq.~(\ref{Eq_warp}), we can rewrite the $H_{eff}$ in polar coordinate as $H_{eff}(k_r)=\lambda_0k_r^2+\lambda_1k_r^3\cos(3\varphi)\tau_z-\mu$. For $\tau=+1$ valley, the $v_f(\varphi)$ can be obtained as
		$v_f(\varphi)=\frac{\partial H_{eff}(k_f(\varphi))}{\partial k_r}=2\lambda_0 k_f(\varphi)+3\lambda_1k_f^2(\varphi)\cos(3\varphi)$. Inserting   $k_f(\varphi)\approx a+b\cos3\varphi$,  the form of $v_f(\varphi)$ is obtained.   We made a plot of $v_f(\varphi)$ with  parameters $\lambda_0=0.5347$ and $\lambda_1=0.0885$ (see the dashed line in Fig.~\ref{fig:figs1}(d)).  Although there is some deviation from the numerical one (in blue), all the symmetry features are captured. To obtain a closer fitting to the numerical results,   one can  expand it to higher order terms, which is not necessary for the purposes of this manuscript. Therefore, we have shown that the trigonal warping effects would result in anisotropic Fermi velocities. The warping term would induce an asymmetry  for the velocities of the incoming and outgoing modes in our scattering matrix method calculations later.

		\subsection{An illustration of valley-polarized states from the Hartree-Fock mean-field approximation}

		More detailed Hartree-Fock mean-field approximation for the moir\'e bands upon Coulomb interaction has been extensively studied in previous works. But for the sake of completeness and illustrate some features of the moir\'e bands upon the valley-polarization, we present the basic formalisms of the valley-polarized states  from the Hartree-Fock approximation with a minimal interacting Hamiltonian:
		\begin{equation}
			H_0=\sum_{\mathbf{k},\tau,s}(\epsilon_{\mathbf{k},\tau}-\mu)\psi^{\dagger}_{\mathbf{k},\tau,s}\psi_{\mathbf{k},\tau,s}+\frac{1}{2A}\sum_{\mathbf{q}}V_{\mathbf{q}}:\rho_{\mathbf{q}}\rho_{-\mathbf{q}}:.
		\end{equation}
		Here, $A$ is the sample area, the density operator, $s$ denotes the spin index, $\mu$ is the chemical potential for the single-particle moir\'e band, $\rho_{\mathbf{q}}=\sum_{\mathbf{k},\mathbf{k'},\tau,s}\braket{c_{\mathbf{k},\tau,s}|e^{i\mathbf{q}\cdot \mathbf{r}}|c_{\mathbf{k'},\tau,s}}c^{\dagger}_{\mathbf{k},\tau}c_{\mathbf{k'},\tau,s}$. Without loss of generality, we focus on the first valence moir\'e band and the Coulomb interaction is projected to this moir\'e band \cite{Fengcheng2020}. The singlet-particle wave function can be decomposed into the plane wave basis as $\ket{c_{\mathbf{k},\tau,s}}=\frac{1}{\sqrt{A}}\sum_{\mathbf{G}}a_{\mathbf{k}+\mathbf{G},\tau, s}e^{i(\mathbf{k}+\mathbf{G})\cdot\mathbf{r}}$ with $\mathbf{G}$ being the moir\'e reciprocal lattice vector. 

		The Hartree-Fock mean-field Hamiltonian with a spin- and valley-polarized ground state  can be written as:
		\begin{equation}
			H^{HF}_{0}\approx \sum_{\mathbf{k},\tau}E_{\mathbf{k}, ,\tau,s}\psi^{\dagger}_{\mathbf{k},\tau,s}\psi_{\mathbf{k},\tau,s},
		\end{equation}
		where
		\begin{eqnarray}
			E_{\mathbf{k},\tau,s}&&=(\epsilon_{\mathbf{k},\tau}-\mu)+\Delta_{\mathbf{k},\tau,s},\\
		\Delta_{\mathbf{k},\tau,s}	&&=\frac{1}{A}\sum_{\mathbf{k'},\tau',s'}V^{\tau\tau'\tau'\tau}_{\mathbf{k}\mathbf{k'}\mathbf{k'}\mathbf{k},ss's's}n_F(E_{\mathbf{k'},\tau',s'})\nonumber
			\\&&-\frac{1}{A}\sum_{\mathbf{k'}}V_{\mathbf{k}\mathbf{k'}\mathbf{k}\mathbf{k'}, ssss}^{\tau\tau\tau\tau}n_{F}(E_{\mathbf{k',\tau,s}}).\label{HF_band}
		\end{eqnarray}
		Here,$n_F$ is the Fermi-Dirac occupation function, the first term of the right-hand of Eq.~(\ref{HF_band}) is the Hartree energy, while the second term is the Fock energy. The Coulomb interaction strength is written as 
		\begin{eqnarray}
			&&V^{\tau\tau'\tau'\tau}_{\mathbf{k_1}\mathbf{k_2}\mathbf{k_3}\mathbf{k_4},ss's's}=\sum_{\mathbf{q}}V_{\mathbf{q}}\braket{c_{\mathbf{k_1},\tau,s}|e^{i\mathbf{q}\cdot\mathbf{r}}|c_{\mathbf{k_4},\tau,s}}\nonumber\\ &&\braket{c_{\mathbf{k_2},\tau',s'}|e^{-i\mathbf{q}\cdot\mathbf{r}}|c_{\mathbf{k_3},\tau',s'}}\nonumber\\
			&&=\sum_{\mathbf{q},\mathbf{G}_1,\mathbf{G}_2,\mathbf{G}_3,\mathbf{G}_4}V_{\mathbf{q}}a^*_{\mathbf{k_1}+\mathbf{G_1},\tau,s}a^*_{\mathbf{k_2}+\mathbf{G_2},\tau',s'}a_{\mathbf{k_3}+\mathbf{G_3},\tau',s'}	\nonumber\\
		 &&a_{\mathbf{k_4}+\mathbf{G_4},\tau,s}\delta_{\mathbf{k_1}+\mathbf{G}_1,\mathbf{q}+\mathbf{k_4}+\mathbf{G_4}}\delta_{\mathbf{k_3}+\mathbf{G}_3,\mathbf{q}+\mathbf{k_2}+\mathbf{G_2}}.
		\end{eqnarray}
		Hence, the Coulomb interactions in the Hartree term and Fock term are written as
		\begin{eqnarray}
			&&V^{\tau\tau'\tau'\tau}_{\mathbf{k}\mathbf{k'}\mathbf{k'}\mathbf{k},ss's's}=\sum_{\mathbf{q},\mathbf{G}_1,\mathbf{G}_2,\mathbf{G}_3,\mathbf{G}_4}V_{\mathbf{q}}a^*_{\mathbf{k}+\mathbf{G_1},\tau,s}a^*_{\mathbf{k'}+\mathbf{G_2},\tau',s'}\nonumber\\
			&&a_{\mathbf{k'}+\mathbf{G_3},\tau',s'}a_{\mathbf{k}+\mathbf{G_4},\tau,s}\delta_{\mathbf{G}_1,\mathbf{q}+\mathbf{G_4}}\delta_{\mathbf{G}_3,\mathbf{q}+\mathbf{G_2}},\\
			&&V_{\mathbf{k}\mathbf{k'}\mathbf{k}\mathbf{k'}, ssss}^{\tau\tau\tau\tau}=\sum_{\mathbf{q},\mathbf{G}_1,\mathbf{G}_2,\mathbf{G}_3,\mathbf{G}_4}V_{\mathbf{q}}a^*_{\mathbf{k}+\mathbf{G_1},\tau,s}a^*_{\mathbf{k'}+\mathbf{G_2},\tau,s}a_{\mathbf{k}+\mathbf{G_3},\tau,s}	\nonumber\\
		 &&a_{\mathbf{k'}+\mathbf{G_4},\tau,s}\delta_{\mathbf{k}+\mathbf{G}_1,\mathbf{q}+\mathbf{k'}+\mathbf{G_4}}\delta_{\mathbf{k}+\mathbf{G}_3,\mathbf{q}+\mathbf{k'}+\mathbf{G_2}}.
		\end{eqnarray} 
		One can solve  Eq.~(\ref{HF_band}) in a self-consistent way. Note that the doubly counted interacting energy should be subtracted after the mean-field approximation, which is given by
		\begin{eqnarray}
			E_0&&=\frac{1}{2A}\sum_{\mathbf{k},\mathbf{k'},\tau,\tau',s,s'}[V^{\tau\tau'\tau'\tau}_{\mathbf{k}\mathbf{k'}\mathbf{k'}\mathbf{k},ss's's}-V^{\tau\tau'\tau'\tau}_{\mathbf{k}\mathbf{k'}\mathbf{k}\mathbf{k'},ss's's}\nonumber\\
			&&\delta_{\tau,\tau'}\delta_{s,s'}]n_F(E_{\mathbf{k},\tau,s})n_F(E_{\mathbf{k'},\tau',s'}).
		\end{eqnarray}
		
		Depending on the filling and the Coulomb interaction strength,  various  time-reversal breaking states with valley-polarization which satisfy the self-consistent Hartree-Fock equation can be obtained, including  i) The fully valley-polarized, spin-unpolarized  insulating or semi-metallic state.
		This state can appear when the Coulomb interaction strength is strong and the filling factor is near some integer fillings;
		ii) The partially valley-polarized, spin-unpolarized metallic states,  which can appear when the Coulomb interactions are weaker \cite{Zaletel2020}; and 
		iii) The valley-polarized, spin-polarized states.  A schematic illustration of the valley-polarized states is presented in the main text Fig.~1. As the focus of this work is the valley-polarized state, we neglect the spin polarization and rewrite the mean-field Hamiltonian as
		\begin{equation}
			H^{HF}_{0}\approx \sum_{\mathbf{k},\tau}\psi^{\dagger}_{\mathbf{k},\tau}[(\tilde{\epsilon}_{\mathbf{k},\tau}-\mu)+\Delta_{\text{vp},\mathbf{k}}\tau_z]\psi_{\mathbf{k},\tau},
		\end{equation}
		where $ \tilde{\epsilon}_{\mathbf{k},\tau}=\epsilon_{\mathbf{k},\tau}+(\Delta_{\mathbf{k},\tau}+\Delta_{\mathbf{k},-\tau})/2$ and $\Delta_{\text{vp},\mathbf{k}}=(\Delta_{\mathbf{k},\tau}-\Delta_{\mathbf{k},-\tau})/2$, and  $\Delta_{\text{vp}}\tau_z$ is the valley-polarized order parameter.  Note that here we are directly taking the valley-polarized state as the ansatz state.  It is actually difficult to determine which state is more energetically favorable based on Hartree-Fock consideration alone. Especially when the correlated state appears at the weak-link region,  the coupling with the superconducting regions can also be important.


		\section{Analytical calculations of the  current-phase relation using the scattering matrix method}

		In the main text, we have used  scattering matrix method to show that the studied MATBG Josephson junction is a $\varphi_0$-JJ and find that the valley polarization  and  warping effects are crucial in giving rise to the observed $\varphi_0$-JJ. We  present the corresponding details  in this Appendix section.
		\subsection{Model Hamiltonian}
		To gain some insight into the crucial features of the junction, we  first look at  the limit of $\Delta_{\text{s}}, \Delta_{\text{vp}}\ll  \mu$, i.e.,  the bandwidth is the biggest energy scale. In this case, we can linearize the momentum near Fermi energy for a fixed transverse momentum $k_y$ and obtain a low-energy effective model as
		\begin{equation}
			H=\frac{1}{2}\sum_{\tau\alpha}\sum_{k_y}\int dx\Psi^{\dagger}_{k_y,\tau\alpha}(x)\hat{H}_{k_y,\tau\alpha}(x)\Psi_{k_y,\tau\alpha}(x).
		\end{equation}
		Here, $\tau\pm$ labels the  $\pm K$ valley, $\alpha=+/-$ labels the incoming/outgoing normal states near Fermi energy,  $\Psi_{k_y,\tau\alpha}=(\psi_{k_y, \tau\alpha}(x),\psi^{\dagger}_{k_y, -\tau,-\alpha}(x))^{T}$ denotes the  Nambu basis, and
		\begin{widetext} 
		\begin{equation}
			\hat{H}_{\tau\alpha}(x)=\begin{pmatrix}
				-i\alpha \hbar v_{f,\tau\alpha}(k_y, x)\partial_x+\Delta_{\text{vp}}(x)\tau&\Delta_{\text{s}}(x)\\
				\Delta_{\text{s}}(x)&i\alpha \hbar v_{f,-\tau-\alpha}(k_y,x)\partial_x+\Delta_{\text{vp}}(x)\tau
			\end{pmatrix}\label{Hamil_eff}
		\end{equation}
	\end{widetext}
		with the pairing potential $\Delta_{\text{s}}(x)=\Delta_{\text{s}}(e^{i\frac{\phi}{2}}\theta(-x)+e^{-i\frac{\phi}{2}}\theta(x-d))$, valley polarization  $\Delta_{\text{vp}}(x)=\Delta_{\text{vp}}\theta(x)\theta(d-x)$ (note that here, we have assumed a uniform valley polarization for the sake of simplicity), the longitudinal Fermi velocity at a fixed $k_y$ of the superconducting part and junction part is given by  $v_{f,\tau\alpha}(k_y,x)=v_{s,\tau\alpha}(k_y)[\theta(-x)+\theta(x-d)]+v_{vp,\tau\alpha}(k_y)\theta(x)\theta(d-x)$ (for the compact of notations,  we will denote $v_{f,\tau\alpha}(k_y,x)\equiv v_{f,\tau\alpha}(x) $ in the following).  Here, $\phi$ is the phase difference, $d$ is the length of the junction, $\Delta_{\text{vp}}$ is the valley polarization strength, $v_{s,\tau\alpha}$, $v_{f,\tau\alpha}$ are the longitudinal Fermi momentum along the current direction of the superconducting part and the junction part with valley polarization. One can  verify that the whole Hamiltonian $\hat{H}$ (dimension is eight by eight) preserves particle-hole symmetry $P\hat{H}P^{-1}=-\hat{H}$ but breaks time-reversal symmetry: $T\hat{H}T^{-1}\neq\hat{H}$ if $\Delta_{\text{vp}}$ is finite. Here, $\hat{P}=\rho_x\alpha_x\tau_x \hat{K}$, $\hat{T}=\alpha_x\tau_x\hat{K}$, $\hat{K}$ is complex conjugate,  and $\alpha_j$ , $\hat{K}$ is com $\tau_j$, and $\rho_j$ are Pauli matrices defined in $\alpha=+/-$, valley, and particle-hole space, respectively.  
		
		Note that in general $v_{vp,\tau\alpha}\neq v_{vp,-\tau-\alpha}$ due to the breaking of time-reversal symmetry, but  $v_{vp,\tau\alpha}\approx v_{vp,-\tau-\alpha}$ in the limit of $\Delta_{\text{vp}}\ll E_F'$. On the other hand,  the warping term breaking intra-valley time-reversal symmetry could lead to $v_{vp,\tau\alpha}\neq v_{vp,\tau-\alpha}$, which plays a crucial role in giving rise to the $\varphi_0$ junction as shown later.
		
		It is worth noting  that the model Hamiltonian  resembles that for an S/F/S junction if the valley is regarded as a pseudo-spin (flips sign under both time-reversal and inversion operation). As we will show later, the $\pi$ junction, which was commonly explored in S/F/S junctions, can also be stabilized in S/VP/S junctions. But we would emphasize that  the physics system in our case is very different, given that the polarization appears in valley degree of freedom rather than spin. 
		\begin{figure}
			\centering
			\includegraphics[width=0.5\linewidth]{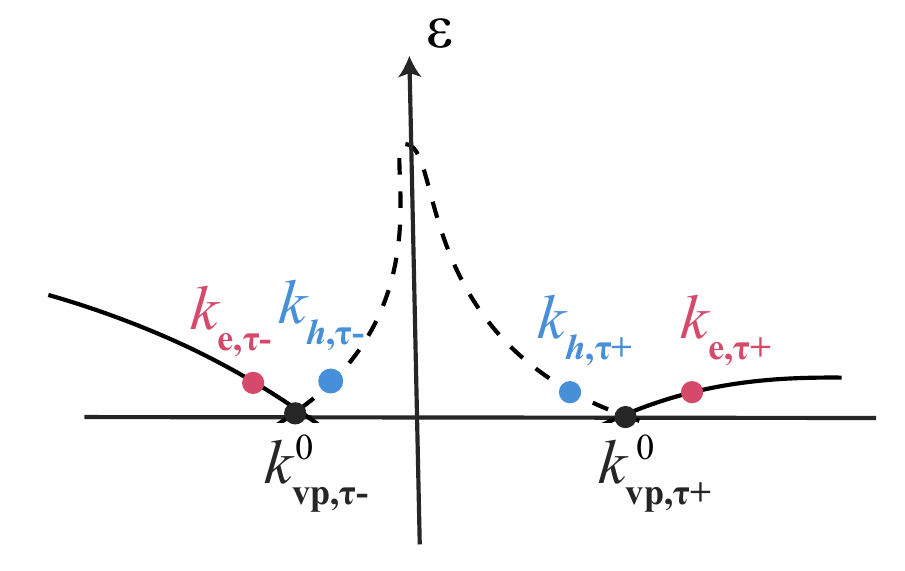}
			\caption{ A schematic plot of the wave vectors $k_{e(h),\tau\alpha}$ on the quasi-particle excitation. The electron- and hole-like quasi-particle bands are highlighted with solid line and dashed line respectively. }
			\label{fig:figs7}
		\end{figure}

		\subsection{Scattering states and boundary conditions}
		The scattering states in the S region of the left(L) and right(R) side are obtained as
		\begin{eqnarray}
		&&	\psi^L_{s,\tau\alpha}=
			\begin{pmatrix}
				e^{-i\alpha\beta}\\e^{-i\frac{\phi}{2}}
			\end{pmatrix}	
			e^{i\alpha k^0_{s,\tau\alpha}x+\kappa_{\tau\alpha}x+i k_yy}
			, x\le 0\\
		&&	\psi^R_{s,\tau\alpha}=\begin{pmatrix}
				e^{i\alpha\beta}\\{e^{i\frac{\phi}{2}}}
			\end{pmatrix}e^{i\alpha k^0_{s,\tau\alpha}(x-L)-\kappa_{\tau\alpha}(x-L)+ik_yy}, x\ge d\nonumber\\
		\end{eqnarray}
		with $k^0_{s,\tau\alpha}$ being the Fermi momentum along the longitudinal direction for a fixed $k_y$ and  the definitions
		\begin{align}
			\kappa_{\tau\alpha}&=\frac{\sqrt{\Delta_{\text{s}}^2-\epsilon^2}}{\hbar v_{s,\tau\alpha}}, \\
			\beta&=\begin{cases}
				\text{acos}\frac{\epsilon}{\Delta_{\text{s}}}, \text{ if } \epsilon<\Delta_{\text{s}},\\
				-i\text{acosh}\frac{\epsilon}{\Delta_{\text{s}}}, \text{ if } \epsilon>\Delta_{\text{s}}.
			\end{cases}
		\end{align}
		The in-gap states $\psi^{L/R}_{s,\tau\alpha}$  with $\epsilon \lesssim \Delta_{\text{s}}$ are superpositions of electron and hole with an exponential decay length $\kappa_{\tau\alpha}^{-1}$ into the left/right superconducting regions. One can verify that the states in the  S region possess time-reversal symmetry: $\psi_{s}(\epsilon,-\phi)=\hat{T}\psi_{s}(\epsilon,\phi)$ with $k^0_{s,\tau\alpha}=k^0_{s,-\tau-\alpha}$.
		
		The scattering state in the VP region $(0\le x\le d)$:
		\begin{eqnarray}
			\psi_{vp,e,\tau\alpha}&=&\frac{1}{\sqrt{N_{e,\tau\alpha}}}\begin{pmatrix}
				1\\0
			\end{pmatrix}e^{ik_{e,\tau\alpha}x+ik_yy},\\
			\psi_{vp,h,\tau\alpha}&=&\frac{1}{\sqrt{N_{h,\tau\alpha}}}\begin{pmatrix}
				0\\1
			\end{pmatrix}e^{ik_{h,\tau\alpha}x+ik_yy}.
		\end{eqnarray} 
		Here, $k_{e,\tau\alpha}$ and $k_{h,\tau\alpha}$ are the wave vectors for electron- and hole-like states, respectively [see an illustration in Fig.~\ref{fig:figs7}], and $N_{e(h),\tau\alpha}$ are normalization factors to ensure that the scattering  matrices are unitary. Up to the leading order, $k_{e,\tau\alpha}\approx k^{0}_{e,\tau\alpha}+\delta k_{e,\tau\alpha},k_{h,\tau\alpha}\approx  k^{0}_{h,\tau\alpha}+\delta k_{h,\tau\alpha}$ with $k^{0}_{e,\tau\alpha}=k^{0}_{h,\tau\alpha}=\alpha k_{vp,\tau\alpha}^0$, and
		\begin{align}
			\delta k_{e,\tau\alpha}=\frac{\epsilon-\tau \Delta_{\text{vp}}}{\alpha \hbar v_{vp,\tau\alpha}}, \\
			\delta k_{h,\tau\alpha}=-\frac{\epsilon-\tau \Delta_{\text{vp}}}{\alpha \hbar v_{vp,-\tau-\alpha}}.
		\end{align}
		Here, $\psi_{vp,e,\tau+}$, $\psi_{vp,h,\tau-}$ are the states moving in the $+x$ direction, while $\psi_{vp,e,\tau-}$, $\psi_{vp,h,\tau+}$ are the states moving in the $-x$ direction.  The particle-hole symmetry requires $\psi_{vp,h}(-\epsilon)=\hat{P}\psi_{vp,e}(\epsilon)$ so that  $\delta k_{h,\tau\alpha}(-\epsilon)=-\delta k_{e,-\tau-\alpha}(\epsilon)$.  The factors $\sqrt{N_{e,\tau\alpha}}$ and $\sqrt{N_{h,\tau\alpha}}$ are to ensure that the scattering matrix is unitary.

		As $\hat{H}$ is block-diagonalized in the valley space with  $[\tau_z,\hat{H}]=0$, we can solve the scattering matrix for $\tau=+$ and $\tau=-$ separately.   We also assume that the transverse momentum $k_y$ is conserved during the scatterings.
		The boundary conditions at $x=0$ and $x=d$  are:
		\begin{widetext}
		\begin{align}
			&a\psi^L_{s,\tau+}(x=0)+b\psi^L_{s,\tau-}(x=0)=c^+_e\psi_{vp,e,\tau+}(x=0)+c^-_e\psi_{vp,e,\tau-}(x=0)\nonumber\\&+c^+_h\psi_{vp,h,\tau+}(x=0)+c^-_h\psi_{vp,h,\tau-}(x=0),\label{boundary1}\\
			&a'\psi^R_{s,\tau+}(x=d)+b'\psi^R_{s,\tau-}(x=d)=c^+_e\psi_{vp,e,\tau+}(x=d)+c^-_e\psi_{vp,e,\tau-}(x=d)\nonumber\\&+c^+_h\psi_{vp,h,\tau+}(x=d)+c^-_h\psi_{vp,h,\tau-}(x=d),\label{boundary2}\\
			&a v_{s,\tau+}\psi^L_{s,\tau+}(x=0)+bv_{s,\tau-}\psi^L_{s,\tau-}(x=0)=v_{vp,\tau+}c^+_e\psi_{vp,e,\tau+}(x=0)-v_{vp,\tau-}c^-_e\psi_{vp,e,\tau-}(x=0)\nonumber\\&+v_{vp,-\tau-}c^+_h\psi_{vp,h,\tau+}(x=0)-v_{vp,-\tau+}c^-_h\psi_{vp,h,\tau-}(x=0),\label{boundary3}\\
			&a'v_{s,\tau+}k_{0,\tau+}\psi^R_{s,\tau+}(x=d)-b'v_{s,\tau-}\psi^R_{s,\tau-}(x=d)=v_{vp,\tau+}c^+_e\psi_{vp,e,\tau+}(x=d)-v_{vp,\tau-}c^-_e\psi_{vp,e,\tau-}(x=d)\nonumber\\&+v_{vp,-\tau-}c^+_h\psi_{vp,h,\tau+}(x=d)-v_{vp,-\tau+}c^-_h\psi_{vp,h,\tau-}(x=d).\label{boundary4}
		\end{align}
	\end{widetext} 
		Here, Eqs.~(\ref{boundary1}) and (\ref{boundary2}) are obtained from the continuity of wavefunction, while Eqs.~(\ref{boundary3}) and (\ref{boundary4}) are obtained from the conservation of particle current, which for each state is given by $\text{Im}(\braket{\psi|\frac{\partial H_{\tau\alpha}(\mathbf{r})}{\partial p_x}|\psi})=\braket{\psi|\text{diag}(\alpha v_{f,\tau\alpha},-\alpha v_{f,-\tau-\alpha})|\psi}$.
		

		\subsection{Andreev bound states in the case without normal reflections}
		We now solve the Andreev bound states using the scattering matrix method \cite{Beenakker1991,Beenakker1992}.  First, we need to work out the scattering matrices. We can define
		\begin{align}
			a(L)=a, b(L)=b, a(R)=a', b(R)=b';\label{def1}\\
			c_{e}^{\dagger}(L)=c_{e}^+, c_e^{-}(L)=c_{e}^-, c_{h}^+(L)=c_{h}^+, c_h^-(L)=c_{h}^-;\\
			c_{e}^{\dagger}(R)=c_{e}^+e^{i(k^0_{e,\tau+}+\delta k_{e,\tau+})d}, c_e^{-}(R)=c_{e}^-e^{i(k^0_{e,\tau-}+\delta k_{e,\tau-})d},\\ c_{h}^+(R)=c_{h}^+e^{i(k^0_{h,\tau+}+\delta k_{h,\tau+})d}, c_h^-(R)=c_{h}^-e^{i(k^0_{h,\tau-}+\delta k_{h,\tau-})d}\label{def4}.
		\end{align}
		The scattering  matrices in the scattering method are defined as
		\begin{equation}
			\begin{pmatrix}
				c_e^+(L)\\c_h^{-}(L)
			\end{pmatrix}=S_L\begin{pmatrix}
				c_e^-(L)\\c_h^+(L)
			\end{pmatrix},  \begin{pmatrix}
				c_e^-(R)\\c_h^{+}(R)
			\end{pmatrix}=S_R\begin{pmatrix}
				c_e^+(R)\\c_h^-(R)
			\end{pmatrix},
		\end{equation}
		and the transition matrices are defined as
		\begin{equation}
			\begin{pmatrix}
				c^+_e(R)\\c_h^-(R)	\end{pmatrix}=T_{RL}\begin{pmatrix}
				c^+_e(L)\\c_h^-(L)
			\end{pmatrix}, 	\begin{pmatrix}
				c^-_e(L)\\c_h^+(L)	\end{pmatrix}=T_{LR}\begin{pmatrix}
				c^-_e(R)\\c_h^+(R)
			\end{pmatrix}.\label{tans}
		\end{equation}

		According to the scattering matrix method \cite{Beenakker1991,Beenakker1992}, the energies of Andreev bound states are given by
		\begin{equation}
			\text{Det}[1-T_{LR}S_RT_{RL}S_L]=0.\label{Eq_Anreev}
		\end{equation}
		The transmission matrices can be directly obtained according to the definitions Eqs.~(\ref{def1}) to (\ref{def4}) and  Eq.~\ref{tans} as
		\begin{eqnarray}
			T_{RL}=\begin{pmatrix}
				e^{i(k^0_{e,\tau+}+\delta k_{e,\tau+})d}&0\\
				0&e^{i(k^0_{h, \tau-}+\delta k_{h,\tau-})d}
			\end{pmatrix},\\
			T_{LR}=\begin{pmatrix}
				e^{-i(k^0_{e, \tau-}+\delta k_{e,\tau-})d}&0\\
				0&e^{-i(k^0_{h,\tau+}+\delta k_{h,\tau+})d}
			\end{pmatrix}.	
		\end{eqnarray}

		The form of scattering matrix $S_{L(R)}$ would depend on the interface at $x=0$ and $x=d$. Let us first consider the case without chemical potential difference between the superconducting region and valley polarized region, i.e., $\mu=\mu'$.  In this case, $v_{vp,\tau\pm}=v_{s,\tau\pm}$, the factors in the scattering states can be simply taken as $N_{e,\tau\alpha}=N_{h,\tau\alpha}=1$.  Using the definitions Eqs.~(\ref{def1}) to (\ref{def4}) and the boundary conditions  Eqs.~(\ref{boundary1}) to (\ref{boundary4}), one can easily obtain
		\begin{eqnarray}
			S_{L(R)}=\begin{pmatrix}
				0&e^{\pm i\frac{\phi}{2}-i\beta}\\
				e^{\mp i\frac{\phi}{2}-i\beta}&0
			\end{pmatrix}.\label{scatter1}
		\end{eqnarray}
		Here,  $ \beta=	\text{acos}\frac{\epsilon_{\tau}}{\Delta_{\text{s}}}$ for in-gap Andreev bound states, and only Andreev reflections in the scattering matrix are finite due to the absence of momentum mismatches. By inserting the scattering matrix back to Eq.~(\ref{Eq_Anreev}), we find that the energies of Andreev bound states are given by
		\begin{equation}
			\cos(2\beta-\frac{2(\epsilon_{\tau}-\tau\Delta_{\text{vp}})d}{\hbar \overline{v}_{vp}})=\cos(\phi+\frac{(\epsilon_{\tau}-\tau\Delta_{\text{vp}})d}{\tau \hbar \delta v_{vp}})\label{bound1},
		\end{equation}
		where
		\begin{eqnarray}
			\bar{v}_{vp}&&=\frac{4}{\sum_{\tau\alpha}v_{vp,\tau\alpha}^{-1}},\\ \delta \bar{v}_{vp}&&=\frac{2}{v_{vp,++}^{-1}+v_{vp,--}^{-1}-v_{vp,+-}^{-1}-v_{vp,-+}^{-1}}.
		\end{eqnarray}
		
		We can further define two energy scales: one is the Thouless energy $E_{T}=\hbar \bar{v}_{vp}/d$, and  the other one is $E_{A}=\hbar \delta \bar{v}_{vp}/d$, which reflects the intra-valley asymmetry induced by the warping term. Then Eq.~\ref{bound1} is rewritten as
		\begin{equation}	\cos(2\beta-\frac{2(\epsilon_{\tau}-\tau\Delta_{\text{vp}})}{E_T})=\cos(\phi+\frac{\epsilon_{\tau}-\tau\Delta_{\text{vp}}}{\tau E_A}).
		\end{equation}
		It clearly shows that the phase $\phi$ is shifted as $\tilde{\phi}=\phi-\varphi_0$ with
		\begin{equation}
		\varphi_0=\Delta_{\text{vp}}/E_A	
		\end{equation} 
	due to the combination of valley polarization and warping effects. As we will see later,  $\varphi_0$ would manifest as the phase shift in a current-phase relation, which would result in the so-called $\varphi_0$ junction. In the short junction limit,  $\epsilon \ll E_A, E_T$,  we can actually obtain the energies of the bound states:
		\begin{equation}
			\epsilon_\tau=\Delta_{\text{s}}\sqrt{1-\sin^2(\frac{\tilde{\phi}}{2}- \frac{\tau\Delta_{\text{vp}}}{E_T})}.\label{bound_energy}
		\end{equation}

		\subsection{The angular dependence of the $\varphi_0$ phase shift}
		Next, we briefly comment on the angular dependence of the  $\varphi_0$ phase shift. It is important to note that the magnitude  of $\varphi_0$ in general would depend on the angle $\theta$ between the current direction and lattice orientation. As an illustration, we can evaluate the $\varphi_0$ phase shift with the approximated angular dependence of the Fermi velocity presented in Sec. I: $v_f(\theta)=\alpha_0+\beta_0\tau\cos3\theta$, where $\tau$ is the valley index, $\alpha_0$ captures the isotropic part, and $\beta_0$ captures the anisotropic part of the Fermi velocity. It is straightforward to obtain $\varphi_0$ according to the relation between $E_A$ and $v_f(\theta)$, which gives 
		\begin{equation}
			\varphi_0(\theta)=\frac{2\beta_0 \cos(3\theta) d\Delta_{vp}}{\hbar (\beta_0^2\cos^23\theta-\alpha^2)}.
		\end{equation}
		Therefore, it can be seen that the $\varphi_0$ phase shift would exhibit a three-fold symmetry: $\varphi_0(\theta)=\varphi_0(\theta+\frac{2\pi}{3})$, depending on the lattice orientation and the current direction.

		\subsection{Andreev bound states in the case with normal reflections}
		In general, the chemical potential between the superconducting region and valley-polarized region is different with $\mu\neq \mu'$. To see the effects of such a difference in chemical potential,  we solved the scattering matrices  in the same way   as
		\begin{eqnarray}
			S_{L(R)}&=&\begin{pmatrix}
				r_N&r_{A}e^{\pm i\frac{\phi}{2}-i\beta}\\
				r_Ae^{\mp i\frac{\phi}{2}-i\beta}&r_N
			\end{pmatrix}\label{scatter2}
		\end{eqnarray}
		with
		\begin{widetext}
		\begin{eqnarray}
			r_{A}&=&e^{i\beta}X^{-1}(v_{vp,\tau+}+v_{vp,\tau-})(v_{s,\tau+}+v_{s,\tau-}),\\
			r_N&=&2iX^{-1}\sin\beta \sqrt{(v_{vp,\tau+}+v_{s,\tau-})(v_{vp,\tau-}+v_{s,\tau+})(v_{vp,\tau+}-v_{s,\tau+})(v_{vp,\tau-}-v_{s,\tau-})},\\
			X&=&e^{i\beta}(v_{vp,\tau+}+v_{s,\tau-})(v_{vp,\tau-}+v_{s,\tau+})-e^{-i\beta}(v_{vp,\tau+}-v_{s,\tau+})(v_{vp,\tau-}-v_{s,\tau-}).
		\end{eqnarray}
	\end{widetext}
		Here, $r_A$, $r_{N}$ are coefficients for  Andreev reflections and normal reflections. Note that we have used $N_{e,\tau+}=N_{h,\tau+}=\sqrt{(v_{vp,\tau+}-v_{s,\tau+})(v_{vp,\tau+}+v_{s,\tau-})}$, $N_{e,\tau-}=N_{h,\tau-}=\sqrt{(v_{vp,\tau-}-v_{s,\tau-})(v_{vp,\tau-}+v_{s,\tau+})}$ .  One can verify that the scattering matrix is unitary with $|r_A|^2+|r_N|^2=1$ for the in-gap bound states with $\epsilon<\Delta$. Evidently, the normal reflections $r_N$ would be finite due to the momentum mismatches, i.e., $v_{vp,\tau\pm}\neq v_{s,\tau\pm}$   induced by the difference in the chemical potential ($\mu\neq \mu'$). It can also be seen that the scattering matrix Eq.~(\ref{scatter2}) would return   to the Eq.~(\ref{scatter1}) if there is no momentum mismatch.
		
		Next, we solve the energies of Andreev bound states in  the case of finite normal reflections. For the compact of notations, we rewrite the scattering matrix as:
		\begin{equation}
			S_{L(R)}=\begin{pmatrix}
				ire^{i\eta}& \sqrt{1-r^2}e^{i\eta}\\
				\sqrt{1-r^2}e^{i\eta}&	ire^{i\eta}
			\end{pmatrix}.
		\end{equation}
		Here, $r=|r_N|$, $\eta=\text{Arg}[X^{-1}]$. By substituting the scattering back to Eq.~(\ref{Eq_Anreev}), we find that the Anreev bound states are given by
		\begin{align}
			&\cos(2\eta+\frac{2(\epsilon-\tau\Delta_{\text{vp}})}{E_T})+r^2\cos(\sum_{\alpha}k^0_{\tau\alpha}d)\nonumber\\&=(1-r^2)\cos(\phi+\frac{\epsilon-\tau\Delta_{\text{vp}}}{\tau E_A})
		\end{align}
		As expected,  the phase shift  $\varphi_0=\frac{\Delta_{\text{vp}}}{E_A}$ would not be affected by the presence of normal reflections. Instead,  the normal reflection would mainly weaken  the magnitude of the supercurrent and thus is not essential for our study.  
		%
		%
		%
		%

		\subsection{Free energy and Josephson currents}
		The free energy of a JJ can be written as
		\begin{equation}
			F=\int d\mathbf{r}\frac{|\Delta_{\text{s}|}^2}{U}-\frac{1}{\beta}\sum_{\epsilon_n}\ln(1+e^{-\beta \epsilon_n}),
		\end{equation}
		where $\epsilon_n$ is the eigenenergies  of the  BdG  Hamiltonian of the Josephson junction, $\beta=1/k_BT$, $U$ is an effective interaction strength.  We neglect the $U$ dependent term which is independent of $\phi$.  One can further subtract a constant normal state free energy $F(\Delta_{\text{s}}=0)$ to avoid the divergence at large energies and would not affect the current-phase relation $I_s(\phi)$ \cite{Beenakker1992}.    The supercurrent through the JJ can be obtained from the free energy with
		\begin{eqnarray}
			I_s(\phi)&&=\frac{2e}{\hbar}\frac{\partial F}{\partial \phi}=\frac{2e}{\hbar}\sum_{\epsilon_n}\frac{1}{e^{\beta\epsilon_n}+1}\frac{\partial \epsilon_n}{\partial \phi}\nonumber\\&&=-\frac{2e}{\hbar}\sum_{\epsilon_n>0}\tanh(\frac{\beta\epsilon_n}{2})\frac{\partial \epsilon_n}{\partial \phi}.\label{current_derive}
		\end{eqnarray}
		Here, $e$ is the charge of an electron.
		One can easily figure out the current units by using $\hbar\approx6.581\times 10^{-13}$ meV$\cdot$s and $e/s\approx1.6\times 10^{-19} $A (A is Ampere), i.e., $2e/\hbar\approx 486$ nA/meV.


		By substituting the bound state energy Eq.~(\ref{bound_energy}) into Eq.~(\ref{current_derive}), and at the high temperature limit $ \Delta_{\text{s}}/k_BT\ll 1$, we obtain Eq.~(7) of the main text:
		\begin{equation}
			I_s(\phi)\approx \frac{e\Delta_\text{s}^2}{2\hbar k_BT}\cos(\frac{2\Delta_{\text{vp}}}{E_T})\sin(\phi-\frac{\Delta_{\text{vp}}}{E_A}).\label{current_ana}
		\end{equation}  

		\subsection{The scattering modes of different transverse momentum $k_y$}
		In the previous sections, we have solved the 1D scattering matrix problem for each mode at a fixed $k_y$. To obtain the total supercurrent through the junction, we need to insert different  longitudinal Fermi momentum $v_{f,\tau\alpha}(k_y)$, and sum over different $k_y$ that are quantized by the finite width.  Unfortunately, we could not do it analytically due to the complicated warping effects.  For the completeness, we still present a brief discussion of the effects of $k_y$ here.  
		
		The total supercurrent through this Josephson junction is given by
		\begin{align}
			I_s(\phi)=\sum_{k_y}I_{s,k_y}(\phi).
		\end{align}
		In the short junction and at  the high temperature limit,  the supercurrent at a phase difference $\phi$: $I_{s,k_y}(\phi)$ carried by each mode can be obtained by replacing $E_T$, $E_A$ in Eq.~(\ref{current_ana}) with the ones calculated from  $v_{f,\tau\alpha}(k_y)$.  As shown in Fig.~\ref{fig:figs1}(d), the value of longitudinal Fermi momentum $v_{f,\tau\alpha}(k_y)$ and its asymmetry near $k_y=0$ are similar so that   the resulting current-phase relation is expected to be similar to Eq.~(\ref{current_ana}) for a small transverse momentum. However, the situation becomes complicated in the large transverse momentum $k_y$. Because of the warping effects, there are multiple scattering modes near Fermi energy for a fixed $k_y$ [see Figs.~\ref{fig:figs1}(c) and \ref{fig:figs1}(d)], which are not captured by Hamiltonian (\ref{Hamil_eff}) that only includes one incoming electron- or hole-dominant mode. Nevertheless, we expect the scattering modes with large momentum to carry less supercurrent and thus in the main text, we find that the 1D scattering Hamiltonian provides a good understanding of our numerical results, in which the current carried by all incoming modes are included.

		\section{  More details for  the  MATBG Josephson junction using the tight-binding method}
		In this section, we present more details about the numerical calculations, including the geometry details, the result in the case of turning off the warping effects, and the result in the case of the weak-link region being a half-filling valley-polarized Chern insulator with a Chern number two.
		\subsection{Model and Geometry details}
		
		\begin{figure}[h]
			\centering
			\includegraphics[width=0.7\linewidth]{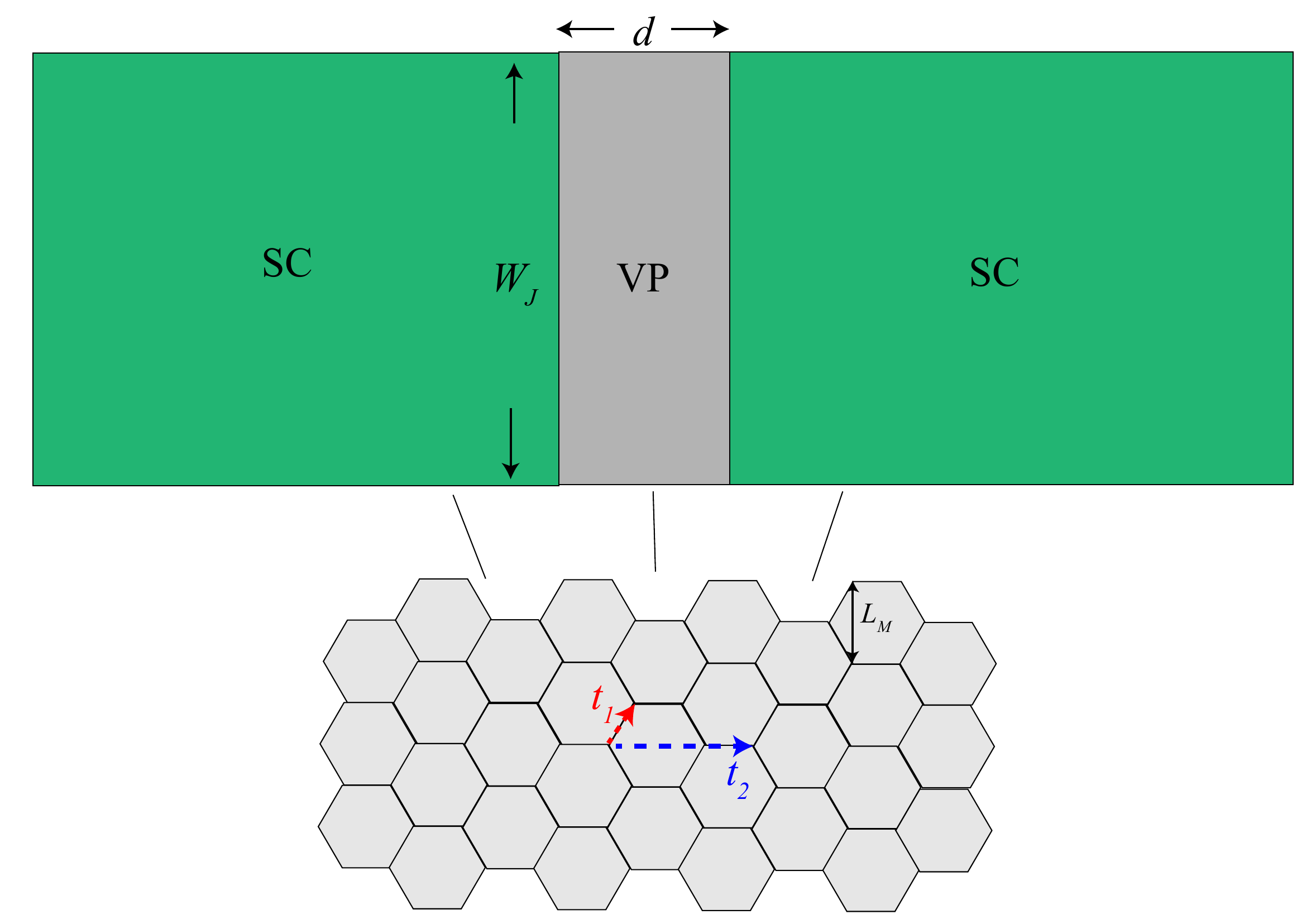}
			\caption{The top panel  presents the junction geometry that is adopted in the  evaluation of Josephson supercurrent through the JJ using  the effective tight-binding model $H$ (Eq.~(\ref{effective_TB}). $d$ and $W_J$ represent the junction length and width, respectively. Here SC, VP represents the region with superconductivity and valley polarization. The bottom panel shows a hexagonal lattice used in the tight-binding model calculation, where $t_1$ denotes the first-nearest hopping , $t_2$  denotes the complex fifth-nearest hopping (resulting in warping term) in $H$, and the  length  of the next nearest bond  is $L_M$.  }
			\label{fig:figs2}
		\end{figure}
		As introduced in the main text, we adopt the following effective tight-binding model  to capture MATBG Josephson junction:
		\begin{eqnarray}\label{effective_TB}
			H_{eff}&&=\sum_{\braket{ij},\xi\sigma}t_1c_{i\xi\sigma}^{\dagger}c_{j\xi\sigma}+\sum_{\braket{ij}',\xi\sigma}t_{2\xi\sigma}c^{\dagger}_{i\xi\sigma}c_{j\xi\sigma}+\text{H.c.}\nonumber\\&&-\sum_{i,\xi}\mu_ic^{\dagger}_{i\xi\sigma}c_{i\xi\sigma}+\sum_{i\in (L,R), \xi}(\Delta_\text{s}e^{i\phi_{L(R)}}c^{\dagger}_{i\xi\uparrow}c^{\dagger}_{i-\xi\downarrow}\text{H.c.})+\nonumber\\
			&&+\sum_{i\in WL,\xi\sigma}\Delta_{\text{vp}}c^{\dagger}_{i\xi\sigma}(\tau_z)_{\xi\xi'}c_{i\xi\sigma}. 
		\end{eqnarray}
		See  the main text for the detailed definitions of the ingredients in Hamiltonian $H_{eff}$.  Here, we depict the adopted geometry of the MATBG Josephson junction in Fig.~\ref{fig:figs2}. The superconducting order parameter $\Delta_{\text{s}}$ and valley-polarized order parameter $\Delta_{\text{vp}}$ are added in the green region and gray region of the top panel of Fig.~\ref{fig:figs2}, respectively. As shown in the bottom panel of Fig.~\ref{fig:figs2}, the lowest moir\'e bands near the charge neutrality are captured by hoppings on the  two-orbital hexagonal lattice in each region, where $t_1$ represents the first-nearest hopping, $t_2$ represents the complex fifth-nearest hopping (giving rise to the warping effects). We note that the minimal tight-binding model proposed in Ref.~\cite{Liang2018} that is used to capture the moir\'e bands up to the lowest hopping is narrower than that from the continuum model shown in Fig.~\ref{fig:figs1}(b). This however would not affect our result as the presence of $\varphi_0$-JJs is determined by the symmetries according to our main text analysis. The key length scales are also highlighted in Fig.~\ref{fig:figs2}. The lattice sites in Fig.~\ref{fig:figs2} label the center of wannier orbitals so that the length of the nearest bonds is the moir\'e lattice constant $L_{M}$. We thus measure the adopted junction length $d$ and $W$ in main text in units of $L_M$.  
		\subsection{Symmetry consideration}
		We now present a symmetry analysis to show why the valley polarization  and the warping effects are crucial for the emergence of  the $\varphi_0$-JJ in MATBG. Without these two ingredients, the system would exhibit time-reversal symmetry which gives $I_{s,\tau}(\phi)\mapsto -I_{s,-\tau}(-\phi)$ and an intravalley inverison symmetry which gives $I_{\text{s},\tau}(\phi)\mapsto -I_{\text{s},\tau}(-\phi)$. As $I_\text{s}(\phi)=\sum_{\tau}I_{\text{s},\tau}(\phi)$, it can be seen that both symmetries would enforce  the total supercurrent to satisfy the condition $I_\text{s}(\phi)=-I_\text{s}(-\phi)$, so that $I_\text{s}(\phi=0)=0$. Therefore, according to our symmetry analysis, the conclusion that valley-polarized state induces $\varphi_0$-JJ is general  as long as time-reversal and intravelley inversion symmetries are broken.  In the below, we present more cases to verify this symmetry analysis.
		\subsection{The case without warping effects}

\begin{figure}[h]
			\centering
			\includegraphics[width=1\linewidth]{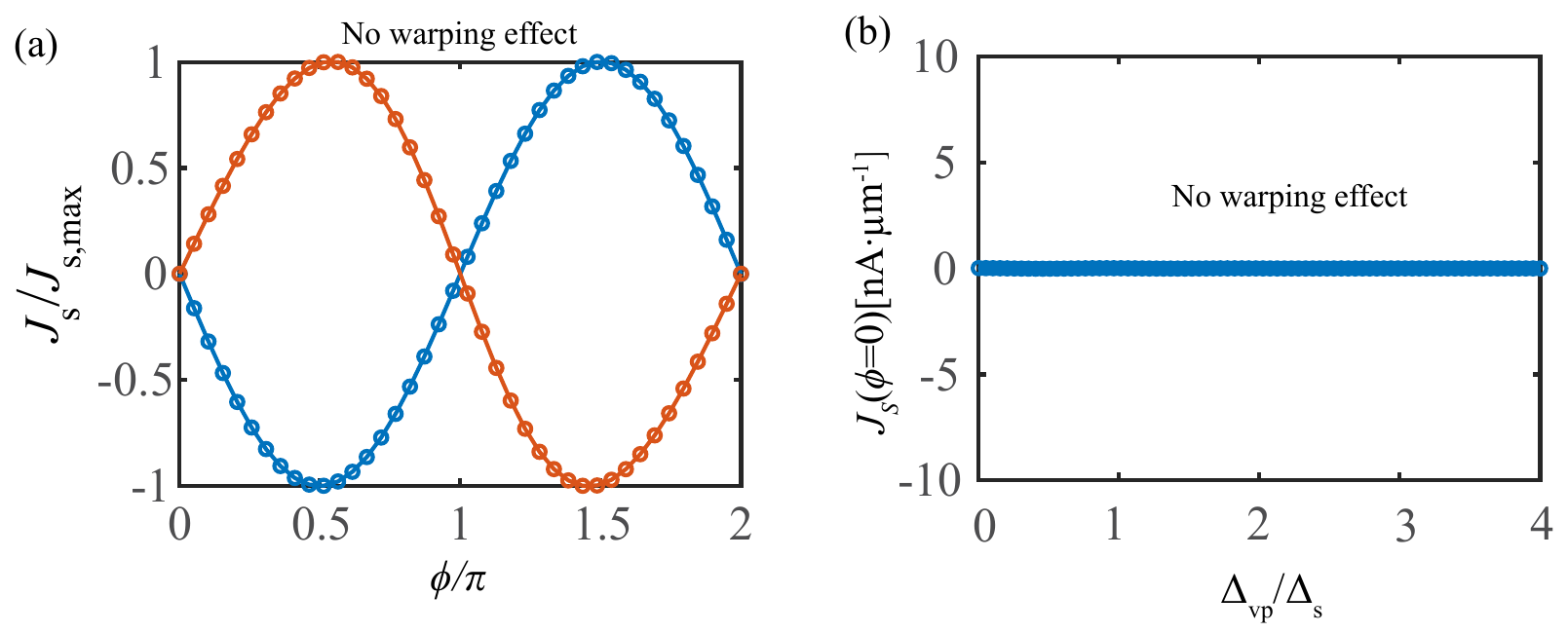}
			\caption{The current-phase relation and vanishing of anomalous Josephson current in the case without warping effects. (a)  Typical curves of Josephson current (normalized by its maximal value) versus Josephson phase difference $\phi$ of the MATBG Josephson junction when the warping term is turned off ($t_{2\xi}=0$), which can only display 0- or $\pi$- junction behavior. (b) The anomalous Josephson current vanishes for various valley polarization strengths where warping effects are not included.  }
			\label{fig:figs3}
		\end{figure}

		In the main text, the warping effects are naturally included in our calculations with the fifth-nearest hopping $t_{2\xi}\neq 0$ (c.f.~\cite{Noah2018,Liang2018}). As discussed in the main text, the warping term would lift the intra-valley inversion symmetry so that the minimal free energy of the junction is not necessary  $0$- or $\pi$-JJ, resulting in a $\varphi_0$-JJ in general. To make a comparison,   we now artificially turn off the warping term. As expected, we find that the junction is restricted to be $0$- or $\pi$-JJ [Fig.~\ref{fig:figs3}(a)]. We consistently find that the anomalous Josephson current, i.e.,$J_s(\phi=0)$, vanishes [Fig.~\ref{fig:figs3}(b)].  It thus clearly shows that the warping effects are crucial for the ground state of MATBG Josephson junction to be $\varphi_0$-JJ, which is in agreement with our symmetry analysis presented in the main text.
		
		\subsection{The case with junction region being valley-polarized Chern insulating states}
		\begin{figure}[h]
			\centering
			\includegraphics[width=0.6\linewidth]{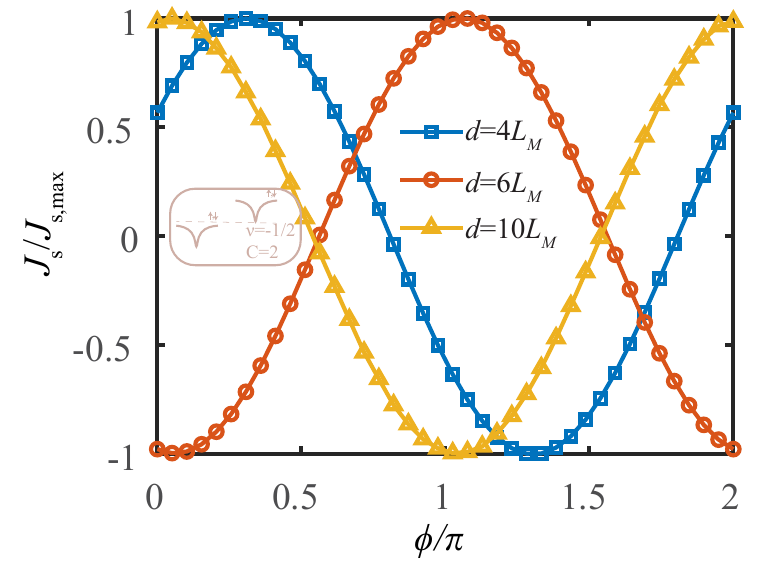}
			\caption{The supercurrent density $J_s$ (normalized by its maximal value) versus the phase difference $\phi$ with  junction length $d=4L_M, 6L_M, 10 L_M$, where the junction region is set to be the valley-polarized Chern insulating states with $C=2$ at half-filling that are illustrated with a schematic inset plot. Here we set the temperature $T=0.05 T_c$.}
			\label{fig:figs4}
		\end{figure}

		It was pointed out in the main text that the mechanism: valley-polarized state mediates unconventional Josephson junction  is quite robust regardless of whether the state is topologically trivial or  nontrivial. In this section,  as a demonstration, we present the calculated current-phase relation [Fig.~\ref{fig:figs4}] by setting the junction region to be half-filling ($\nu=-1/2$) valley-polarized Chern insulating states with Chern number $C=2$ [see a schematic illustration in the inset of Fig.~\ref{fig:figs4}]. One can add a  Haldane term to the tight-binding Hamiltonian (\ref{effective_TB}) in order to make the junction region  topological (c.f.~Ref.~\cite{Diez_Merida2021}). In this case, as shown in Fig.~ \ref{fig:figs4}, the curves of  supercurrent $J_s$ (normalized by its maximal value) versus the phase difference $\phi$ would still display a finite phase shift, i.e., $\sin(\phi-\varphi_0)$,  for various junction lengths $d$. In other words,  the junction would still behave a $\varphi_0$-JJ.  Note that  in the topological case, the edge states that can  mediate some supercurrents may play an additional role. Nevertheless, Fig.~\ref{fig:figs4} clearly shows that our conclusion about the valley polarization  causing $\varphi_0$-JJ is not affected. It is understandable given that time-reversal and intra-valley inversion symmetry are still broken by the valley-polarized Chern bands in this case.

		\section{The magnetic interference for $\varphi_0$-Josephson junctions}

		\subsection{The magnetic interference of a uniform $\varphi_0$ Josephson junction-standard Fraunhofer pattern}
		In this section, we show that the magnetic interference of a uniform $\varphi_0$ Josephson junction should be  the standard Fraunhofer pattern. The gauge invariant phase difference across an extended junction is
		\begin{equation}
			\gamma=\tilde{\phi}+\frac{2e}{\hbar}\int \mathbf{A\cdot}d\mathbf{l}
		\end{equation}
		with $\tilde{\phi}=\phi-\varphi_0$, $\varphi_0$ denoting the phase shift in $\varphi_0$-junction. For an out of plane magnetic field, the gauge can be chosen as $\mathbf{A}=(-By,0)$. 
		The Josephson current is given by
		\begin{equation}
			I_s=\int dy \bm{j}(\mathbf{r},\tilde{\phi}-\frac{2e}{\hbar}\int_{-d/2}^{d/2} dx By).
		\end{equation} 
		
		Assuming that  the current follows the simplest $\sin (\phi-\varphi_0)$ feature, we will obtain
		\begin{eqnarray}
			I_s&=&\int_{-W_J/2}^{W_J/2}dy j(y)\sin(\tilde{\phi}-\frac{2e}{\hbar}Byd)\nonumber.
		\end{eqnarray}
		Here, we denote the width of the junction to be $W_J$. In the case of a uniform current density $j(x)=j_b$, then
		\begin{equation}
			I_s(\Phi)=\frac{j_b\sin\tilde{\phi} \sin(\pi\Phi/\Phi_0)}{\frac{\pi}{\Phi_0}BL},
		\end{equation}
		where the flux quantum $\Phi_0=h/2e$, $\Phi=B\times W_Jd$.
		Thus,
		\begin{equation}
			I_{c}(\Phi)= I_{c}| \frac{\sin(\pi\Phi/\Phi_0)}{\pi \Phi/\Phi_0}|,
		\end{equation}
		where the critical current  at zero-field is denoted as $I_c=j_0W_J$. Hence, for a uniform $\varphi_0$ JJ, our phenomenological calculation suggests that the critical current as a function of external fields $I_c(\Phi)$ follows the standard Fraunhofer pattern.

		\subsection{Phenomenological theory for the magnetic interference of a $\varphi_1$-$\varphi_2$ Josephson junction}
		
		
		As we have mentioned in the main text, the total current through this junction under external magnetic fields can be written as
		\begin{eqnarray}
			I_s(\phi)&&=\int_{-W_{J1}}^{0}dy j_b\sin(\phi-\varphi_1-\frac{2e}{\hbar}Byd)\nonumber\\&&+\int_{0}^{W_{J2}}dy j_b\sin(\phi-\varphi_2-\frac{2e}{\hbar}Byd).\label{phi_12JJ}
		\end{eqnarray}
		Note that $\varphi_1$ and $\varphi_2$ would be different as $\Delta_{\text{vp}}$  in two  domain walls are different. If $\varphi_1=0$ and $\varphi_2=\pi$, the scenario would reduce to the $0$-$\pi$ JJ studied in Ref.~\cite{Weides2006, Frolov2006, Kemmler2010}, where the Fraunhofer pattern exhibits a dip near the zero magnetic flux due to the cancellation of the supercurrent of the $0$-JJ parts and $\pi$-JJ parts. In our case, the $\varphi_1$ and $\varphi_2$ can be a value ranging from $0$ to $2\pi$ due to the formation of $\varphi_0$-JJ. Hence, we call it  $\varphi_1$-$\varphi_2$ JJ.
		
		The Fraunhofer pattern of the $\varphi_1$-$\varphi_2$ JJ can be obtained from Eq.~(\ref{phi_12JJ}). Specifically, the total supercurrent is written as
		\begin{eqnarray}
			&&I_s(\phi)=\frac{I_{s1}}{2\pi\frac{\Phi_1}{\Phi_0}}(\cos(\phi-\varphi_1)-\cos(\phi-\varphi_1+2\pi\frac{\Phi_1}{\Phi_0}))\nonumber\\
			&&+\frac{I_{s2}}{2\pi\frac{\Phi_2}{\Phi_0}}(\cos(\phi-\varphi_2-2\pi\frac{\Phi_2}{\Phi_0})-\cos(\phi-\varphi_2)).
		\end{eqnarray}
		Here,   $I_{s1}=j_bW_{J1}$ and $I_{s2}=j_bW_{J2}$ denote the current through the two domain walls, respectively, and the magnetic flux though the $j$-th domain wall is $\Phi_j=BW_jd$. For the sake of simplicity, we denote $I_{s1}=\frac{1}{2}(1+\delta)I_s$ , $I_{s2}=\frac{1}{2}(1-\delta)I_s$, $\Phi_1=\frac{1}{2}(1+\delta)\Phi$, and $\Phi_2=\frac{1}{2}(1-\delta)\Phi$, where $I_s=I_{s1}+I_{s2}$ is the total supercurrent through the junction, and $\Phi=B(W_{J1}+W_{J2})d$ is the total magnetic flux. Using these notations, the 
		\begin{eqnarray}
			I_s(\phi)&&=\frac{I_s}{\frac{2\pi\Phi}{\Phi_0}}[\cos(\phi-\varphi_1)-\cos(\phi-\varphi_2)+\cos(\phi-\varphi_2\nonumber\\
			&&-\frac{\pi(1-\delta)\Phi}{\Phi_0})-\cos(\phi-\varphi_1+\frac{\pi(1+\delta)\Phi}{\Phi_0})]\nonumber\\
			&&=\frac{I_s}{ \frac{\pi\Phi}{\Phi_0}}[\sin(\phi-\frac{\varphi_{+}}{2}+\delta\frac{\pi\Phi}{\Phi_0})\sin(\frac{\varphi_{-}}{2}+\frac{\pi\Phi}{\Phi_0})\nonumber\\&&-\sin(\phi-\frac{\varphi_{+}}{2})\sin(\frac{\varphi_{-}}{2})],
		\end{eqnarray}
		where $\varphi_{\pm}=\varphi_2\pm \varphi_1$.
		
		The critical current $I_c=\text{max}(I(\phi))$, given by the maximal value of $I_s(\phi)$ within $0\le\phi\le 2\pi$. For the 0-0 JJ, one can easily obtain the standard Fraunhofer pattern	$I_c(\Phi)=I_s|\frac{\sin(\pi\Phi/\Phi_0)}{\pi\Phi/\Phi_0}|$. Due to the presence of the asymmetry parameter $\delta$, beyond 0-0 JJ, we  could only find analytical solutions of the critical current in some special cases , such as  in the limit of $\delta=0$:
		\begin{equation}
			I_c(\Phi)=I_s|\frac{\sin(\frac{\varphi_{-}}{2}+\frac{\pi\Phi}{\Phi_0})-\sin(\frac{\varphi_{-}}{2})}{\frac{\pi\Phi}{\Phi_0}}|.
		\end{equation}
		It can be noted that the critical current is always zero if the magnetic flux reaches certain integer flux quantum $\Phi=2n\Phi_0$ ($n$ are finite integers). As there is no node in the Fraunhofer pattern of experiments, we remove these nodes by introducing a finite asymmetric parameter $\delta$. For example,  at  $\Phi=2n\Phi_0$, the critical current becomes $I_s|\frac{\sin(\frac{\varphi_{-}}{2})\sin\delta n\pi}{2n\pi}|$, which could be finite  if $\delta\neq 0$.
		
		It is worth noting that another key feature of the experimentally observed Fraunhofer pattern is  to exhibit the $I_c(\Phi)\neq I_c(-\Phi)$. In the cases of $0$-$0$ JJ and $0$-$\pi$ JJ, we find that the resulting Fraunhofer patterns are always symmetric, regardless of the choice of $\delta$. However, if we consider $\varphi_0$-JJ, where $\varphi_{\pm}$ can take a more generic value rather than $0$ or $\pi$, we find that the resulting Fraunhofer pattern is asymmetric in general. In  Fig.~4 of the main text, we plotted the Fraunhofer pattern for the $0$-$0$ JJ, $0$-$\pi$ JJ, and $\varphi_1$-$\varphi_2$ JJ with $\delta=0.4, \varphi_1=0.2, \varphi_2=\pi+0.8$. The resulting Fraunhofer pattern arising from $\varphi_1$-$\varphi_2$ JJ is quite consistent with that seen in the experiment. Our calculation thus suggests that the presence of  $\varphi_0$-JJs could  provide a plausible explanation for such highly unconventional Fraunhofer patterns.

		Here we further emphasize how the features of the unconventional  pattern shown in the main text Fig.4 are related to the $\varphi_1$-$\varphi_2$ JJ model, especially the model parameters   $\varphi_1$, $\varphi_2$,  $\delta$:
		(i)The unconventional Fraunhofer pattern (red line), $|I_c(\Phi)\neq I_c(-\Phi)|$, which would indicate the time-reversal breaking. (ii) The unconventional Fraunhofer pattern exhibits a local minimal around zero flux. As a result, the central peak is shifted to a finite flux. It is sharply different from the conventional Fraunhofer pattern (gray line) which exhibits a maximal peak around zero flux. In the $\varphi_1$-$\varphi_2$ JJ, the local minimal appears around zero flux when the difference between $\varphi_1$ and $\varphi_2$ exceeds $\pi$, which results in a cancellation of supercurrent through $\varphi_1$ and $\varphi_2$  junction part.
		(iii)The unconventional Fraunhofer pattern exhibits non-vanishing nodes at finite integer flux. According to the expression $I_c(\Phi=2n\Phi_0)$, the non-vanishing nodes at $\Phi=2n\Phi_0$  indicates$\varphi_1$ and $\varphi_2$ are different, $W_{J1}$ and $W_{J2}$ are different as $\delta\neq 0$.

		
		\subsection{Detection of $\varphi_0$-JJ with a superconducting MATBG SQUID}
		\begin{figure}[h]
			\centering
			\includegraphics[width=1\linewidth]{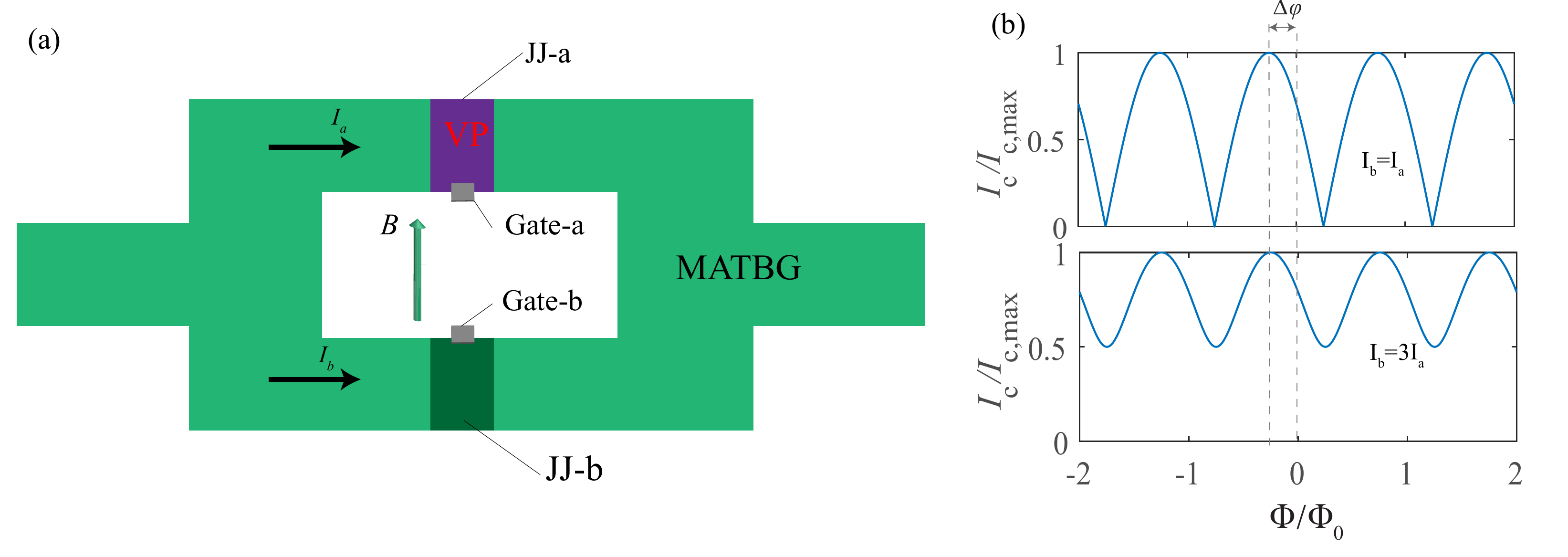}
			\caption{(a) A schematic plot of a MATBG SQUID. Here, JJ-a represents a junction region controlled by gate-a, while JJ-b represents the other junction region controlled by gate-b. $B$ is an out of plane magnetic field. (b) The critical supercurrent (normalized by its maximal value) as a function of the magnetic flux $\Phi$ in the case of $I_b=I_a$ (top panel), $I_b=3I_a$ (bottom panel). $I_a$, $I_b$ are the supercurrent through JJ-a, JJ-b, respectively.  The phase shift of the interference pattern $\Delta\varphi$ is highlighted. }
			\label{fig:figs6}
		\end{figure}
		
		The  SQUID  can be used   to identify the $\varphi_0$-JJ behavior in the experiments \cite{Szombati2016}. For the sake of completeness, as shown in Fig.~\ref{fig:figs6}(a), here we propose a MATBG SQUID geometry to detect the $\varphi_0$-JJ predicted by our theory. In this geometry, there are two weak-linked junction regions that are achieved  by local gates-a,b. Without loss of generality, we consider one is the $\varphi_0-$JJ with the junction gated into valley-polarized states (JJ-a), while the other one is the conventional JJ (JJ-b).
		
		The total supercurrent through the SQUID under  magnetic fields is written as
		\begin{equation}
			I_s=I_b\sin(\phi_b)+I_a\sin(\phi_a-\varphi_0)
		\end{equation}
		with
		\begin{equation}
			\varphi_b-\varphi_a=2\pi\Phi/\Phi_0.
		\end{equation}
		Here, $I_a$($\varphi_a$) and $I_b$($\varphi_b$) are the supercurrent (phase difference) across the JJ-a, JJ-b, respectively. $\Phi$ is the magnetic flux through the SQUID. In a simple case where $I_a=I_b=I_0$, we can obtain the critical current at each magnetic flux as
		\begin{equation}
			I_c=2I_0|\cos(\pi\frac{\Phi}{\Phi_0}+\frac{\varphi_0}{2})|.
		\end{equation}
		Hence, the $\varphi_0$ would cause a phase shift in the SQUID pattern [see the top panel of Fig.~\ref{fig:figs6}(b)], where $\Delta\varphi=\varphi_0/2\pi$. In general, $I_a$ and $I_b$ are not equal. As a illustration, we plot the magnetic interference pattern with $I_b=3I_a$ in Fig.~\ref{fig:figs6}(c). In this case, it can be seen that although  the critical currents no longer vanish at certain magnetic fields, the phase shift does not change. 
		
		Therefore, the proposed MATBG SQUID provides a feasible way to directly measure  the predicted $\varphi_0$ phase shift. Upon finishing our work, we noticed that  the MATBG SQUID geometry has recently been successfully  fabricated in the experiment \cite{Fokert_SQUID_2022}. Our work thus would motive experimentalists to further gate the junction region into valley-polarized states and study the proposed unconventional Josephson effects  in the near future. 
		
		\section{ the $\varphi_0$-JJ beyond conventional pairings}
		\begin{table}[h]
			\caption{Classifications of all possible momentum independent  pairing of  TBG according to the irreducible representations of the $D_3$ symmetry group.}
			\begin{tabular}{cccc}
				\hline\hline
				IRs &\hspace{1 mm} $A_1$  &\hspace{1 mm} $A_2$ &\hspace{1 mm} $E$\\\hline	
				$C_{3z}=\tau_0\otimes e^{-i\frac{\pi}{3}\sigma_z}$ &\hspace{1 mm} $+1$ &\hspace{1 mm} $+1$ &\hspace{1 mm} $+2$\\
				$C_{2y}=\tau_x\otimes i\sigma_y$ &\hspace{1 mm} $+1$ &\hspace{1 mm} $+1$ &\hspace{1 mm} $0$\\\hline
				Spin-singlet &\hspace{1 mm}$\tau_x\otimes i\sigma_y$&\hspace{1 mm} --- &\hspace{1 mm}---\\
				Spin-triplet  &\hspace{1 mm}   $i\tau_y\otimes\sigma_x$         &\hspace{1 mm} ---&\hspace{1 mm} $(i\tau_y\otimes\sigma_z,i\tau_y\otimes\sigma_0)$\\\hline
				\hline
			\end{tabular}
			\label{TableS3}
		\end{table}
		\begin{figure}[h]
			\centering
			\includegraphics[width=0.9\linewidth]{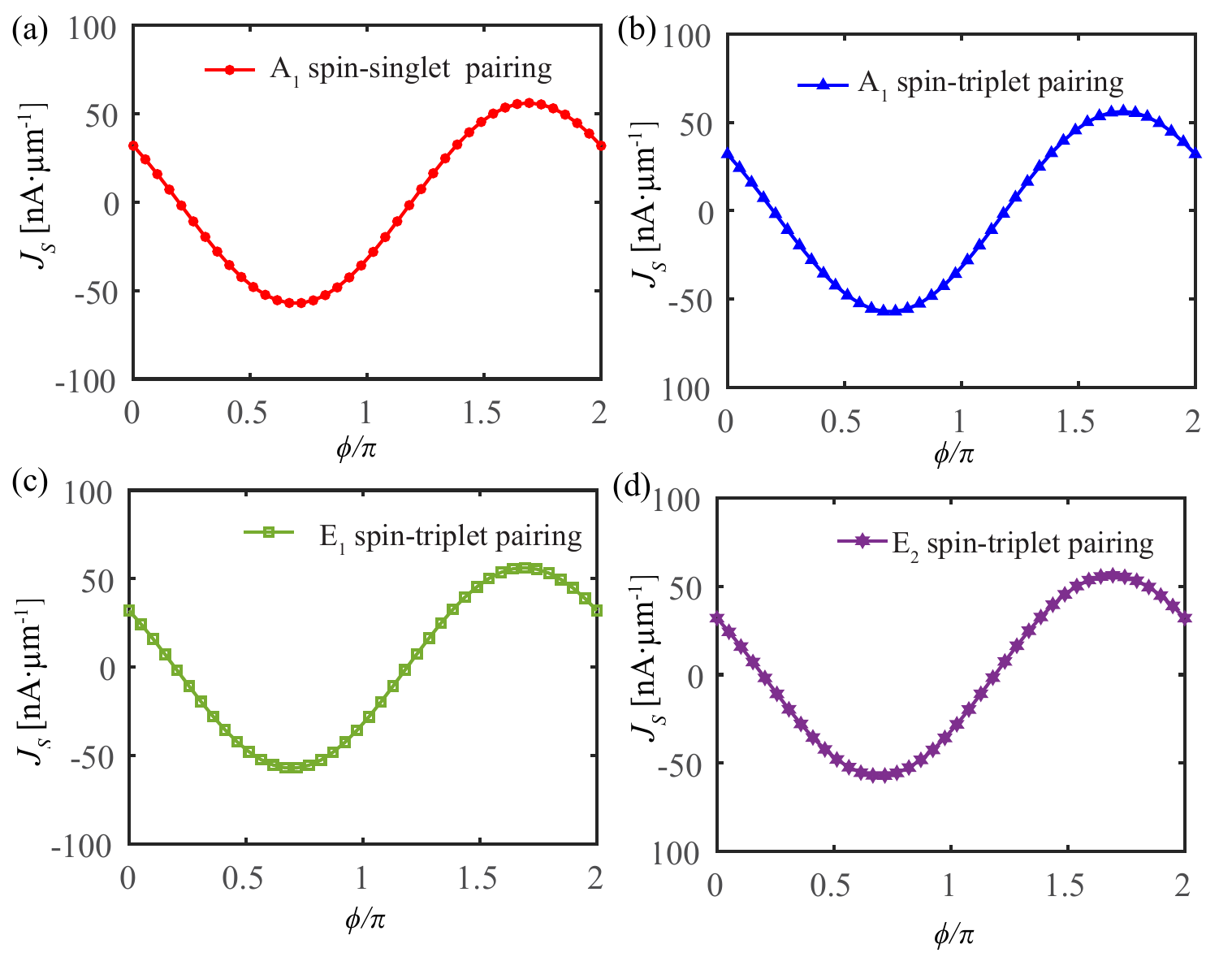}
			\caption{(a)-(d)The supercurrent density $J_s$ (in the unit of nA$\cdot$ $\mu$m$^{-1}$) versus the Josephson phase difference $\phi$ for $A_1$ spin-singlet pairing, $A_1$ spin-triplet pairings, $E_1$ spin-triplet pairing, $E_2$ spin-triplet pairing, respectively. Here we adopt a valley polarization strength $\Delta_{\text{vp}}/\Delta_{\text{s}}=3$ and a temperature $T/T_c=0.3$.}
			\label{fig:figs5}
		\end{figure}
		In the main text, to be specific,  we have adopted the spin-singlet pairing as the pairing order parameter for the superconducting part of the MATBG Josephson junction. In this section, we point out that the $\varphi_0$-JJ can still persist even when the pairing is unconventional, such as various spin-triplet pairings. The pairings can be expanded in the space formed by spin and valley degrees of freedom. We first classify the possible pairings using irreducible representations of the $D_3$ crystal group of MATBG. For simplicity, we focus on all $\mathbf{k}$-independent inter-valley pairings.
		
		Specifically, the generators of $D_3$ point contains a three-fold rotation along the $z$-axis represented by $C_{3z}=\tau_0\otimes e^{-i\frac{\pi}{3}\sigma_z}$, and a two-fold rotation along the $y$-axis represented by $C_{2y}=\tau_x\otimes i\sigma_y$. Here, $\sigma$ and $\tau$ are Pauli matrices defined in spin- and valley-space. Note that $C_{2y}$ would exchange the $K$ and $-K$ valley, while $C_3$ would  not. 
		
		The pairing matrix transforms under the a point group symmetry operation as:
		\begin{equation}
			g\hat{\Delta}_s\mapsto U^{\dagger}(g)\hat{\Delta}_sU^{*}(g), 
		\end{equation} 
		where $\hat{\Delta}_s$ is defined in Nambu basis: $(\psi_{+,\uparrow},\psi_{+,\downarrow},\psi_{-,\uparrow},\psi_{-,\downarrow})^{T}$ with $+/-$  as valley index and $\uparrow/\downarrow$ for spin up/spin down, $U(g)$ is the matrix representation of the generator $g$ in the spin- and valley-space. Note that  $\hat{\Delta}_s=-\hat{\Delta}_{\text{s}}^{T}$ due to the Fermi statistics, and the representation of $\hat{\Delta}_s$  in the valley degree of freedom  is restricted to be $\tau_x$ and $\tau_y$, i.e., inter-valley nature.   All the $\mathbf{k}$-independent inter-valley pairings are summarized in Table \ref{TableS3}.

		There is one inter-valley spin-singlet $A_1$ pairing :  $\Delta_{A_1,s}=\tau_x\otimes i\sigma_y$,, and there are two inter-valley spin-triplet pairings: one one-dimensional spin-triplet $A_1$-pairing  $\Delta_{A_1,t}=i\tau_y\otimes \sigma_x$, and one two-dimensional spin-triplet   $E$-pairing, which we label as $E_1$-pairing and  $E_2$-pairing with $(\Delta_{E,1},\Delta_{E,2})=(i\tau_y\otimes\sigma_z,i\tau_y\otimes\sigma_0)$. Note that the pairings labeled by different irreducible representations do not mix, and   the mixing of $\Delta_{A_1,s}$ and $\Delta_{A_1,t}$ is expected to be neglectable as the spin-orbit coupling in graphene is extremely small. It is also worth noting that the possible nematic pairings can be constructed using the pairing matrices in  the two-dimensional $E$-pairing. 
		
		We can replace the order parameter of the superconducting part   with the above unconventional momentum independent pairings  in the previous tight-binding model calculation and evaluate the supercurrent in the same way. As shown in Figs.~\ref{fig:figs5} (a)-(d), we find that the current-phase relation is unchanged in the cases of various spin-triplet pairings, which thus implies that our result is  not sensitive to the spin configurations of  Cooper pairs of the superconducting part. This observation is understandable as the appearance of $\varphi_0$-JJ is mainly induced by the valley polarization of the junction region.

		\section{A comparison between our model and the $\varphi_0$ JJ model arising from exchange fields and spin-orbit coupling}
		In the main text, we pointed out that the warping term and valley polarization effectively, respectively, play the role of exchange fields and spin-orbit coupling in comparison with  previous $\varphi_0$ JJ models with these terms. Here we illustrate this with more details. 
		To map the trigonal warping term as a spin-orbit coupling, we can look at the k$\cdot$ p normal state Hamiltonian of our system before linearization as shown in Sec. I:
		\begin{equation}
			H_N=\lambda_0 (k_x^2+k_y^2)+\lambda_1 [k_x(k_x^2-3k_y^2)+\Delta_{vp}]\tau_z.
		\end{equation}
		Note that the normal part of the phenomenological model in the main text Eq. (3) is given by linearizing the momentum near Fermi energy. 
		
		If we regard the valley as a pseudospin, the trigonal warping term $\lambda_1 k_x (k_x^2-3k_y^2 ) \tau_z $ indeed can be regarded as a spin-orbit coupling and $\Delta_{vp}\tau_z$ can be regarded as a polarization induced by an exchange field. When we consider the Josephson junction, $k_y$ can be fixed as a good number by setting the current direction to be the $k_x$-direction. For simplicity, we set $k_y=0$ as we considered for the phenomenological model Eq.~(3), and then  $ H_{N,1D}=\lambda_0 k_x^2+[\lambda_1 k_x^3+\Delta_{vp}]\tau_z$.
		
		Next,  we highlight the similarities between our model and the Rashba nanowire model, which can be written as $H_R=k_x^2/2m+\alpha k_x \sigma_y+ h\sigma_y$ with the $h\sigma_y$ denoting the Zeeman coupling of the magnetic field with the spin of the electron in the y-direction, $\alpha$ is the strength of the Rashba spin-orbit coupling. It is known that the Rashba spin-orbit coupling combined with an in-plane 
		spin polarization would results in the $\varphi_0$ JJ.  Our $H_{N,1D}$ model (with $k_y$=0)  is equivalent to replacing  the Rashba spin-orbit coupling model with a cubic warping term, which can be seen by performing a unitary transformation mapping $\tau_z$ to $\tau_y$  in $H_{N,1D}$: $H'_{N,1D}=\lambda_0 k_x^2+\lambda_1(k_x^3+\Delta_{vp})\tau_y$. After this mapping, it is indeed understandable that we can get a $\varphi_0$ JJ. 
		
		However, it should be noted that the underlying physical system in our case is very different, given that the polarization appears in valley degrees of freedom rather than spin. Our proposal relies on the valley-polarized moiré bands, and it does not need to involve spin-orbit coupling or exchange fields.    Only when we restrict the interaction-induced order parameter to the valley polarization $\Delta_{vp}\tau_z$, our model can be mapped to the model with Rashba SOC in some limit. Indeed, the interaction-induced order parameter can be richer.   Another difference we would like to highlight here is that the possessing of both spin and valley  degree of freedom in moiré bands allows much more rich cases with various band alignment. For example,  the spin-valley polarized state  can also appear at very low temperatures in the experiment \cite{Diez_Merida2021}. In this case, the order parameter can be written as
		$\Delta_{vp}\tau_z\sigma_0+\Delta_{sp} \tau_0\otimes\sigma_z$.  Both the spin polarization and valley polarization would contribute to the anomalous phase shift. However, this model cannot be generally mapped to a Rashba spin-orbit coupling model as we would deal with a four by four matrix with two kinds of polarization.
		\end{appendix}
	 %
	 
 
%
%

\end{document}